\documentclass[10pt,journal,compsoc]{IEEEtran}  

\usepackage{tabu}
\usepackage{graphicx}
\usepackage[tight,footnotesize]{subfigure}
\usepackage{tabularx}
\usepackage{listings}
\usepackage{pifont} 
\usepackage{makecell} 
\usepackage{booktabs} 
\usepackage{verbatim}
\usepackage{layouts}
\usepackage{booktabs}
\usepackage{multirow}
\usepackage{url} 

\UseRawInputEncoding 

\newcommand{\xmark}{\ding{55}}
\newcommand{\cmark}{\ding{51}}

\ifCLASSOPTIONcompsoc
  \usepackage[nocompress]{cite}
\else
  \usepackage{cite}
\fi
\ifCLASSINFOpdf
\else
\fi
\begin{document}
\newpage
\onecolumn
\begin{table*}
	\resizebox{0.9\linewidth}{!}{%
		\begin{tabular}{l}		
			\textbf{Copyright © 2022 IEEE}	\\
			\textbf{© 2022 IEEE. Personal use of this material is permitted. Permission from}\\
			\textbf{IEEE must be obtained for all other uses, in any current or future media,}\\
			\textbf{including reprinting/republishing this material for advertising or promotional}	\\
			\textbf{purposes, creating new collective works, for resale or redistribution to servers}\\
			\textbf{or lists, or reuse of any copyrighted component of this work in other works.}\\
			\textbf{}\\  	[1ex] 	           
	\end{tabular}}
	
\end{table*}

\newpage
\twocolumn	
	
%
\title{Containerisation for High Performance Computing Systems: Survey and Prospects}
%
%
%
%
\author{Naweiluo Zhou, Huan Zhou, Dennis Hoppe	
	
	\IEEEcompsocitemizethanks{\IEEEcompsocthanksitem All authors are with High Performance Computing Center Stuttgart (HLRS), University of Stuttgart, Stuttgart, Germany.\protect \\ E-mail: naweiluo.zhou@ieee.org, huan.zhou@hlrs.de, hoppe@hlrs.de
	}

}

\markboth{Journal of Software Engineering,~Vol.~X, No.~X, May~2022 (This is the authors' version)}%
{Shell \MakeLowercase{\textit{et al.}}: Bare Demo of IEEEtran.cls for Computer Society Journals}
%



\IEEEtitleabstractindextext{%
\begin{abstract}
Containers improve the efficiency in application deployment and thus have been widely utilised on Cloud and lately in High Performance Computing (HPC) environments. Containers encapsulate complex programs with their dependencies in isolated environments making applications more compatible and portable. Often HPC systems have higher security levels compared to Cloud systems, which restrict users' ability to customise environments. Therefore, containers on HPC need to include a heavy package of libraries making their size relatively large. These libraries usually are specifically optimised for the hardware, which compromises portability of containers. \textit{Per contra}, a Cloud container has smaller volume and is more portable. Furthermore, containers would benefit from orchestrators that facilitate deployment and management of containers at a large scale. Cloud systems in practice usually incorporate sophisticated container orchestration mechanisms as opposed to HPC systems. Nevertheless, some solutions to enable container orchestration on HPC systems have been proposed in state of the art. This paper gives a survey and taxonomy of efforts in both containerisation and its orchestration strategies on HPC systems. It highlights differences thereof between Cloud and HPC. Lastly, challenges are discussed and the potentials for research and engineering are envisioned. 

\end{abstract}

\begin{IEEEkeywords}
HPC, Container, Orchestration, Resource Management, Job Scheduling, Cloud Computing, AI 
\end{IEEEkeywords}}

\maketitle

\IEEEdisplaynontitleabstractindextext

%
\IEEEpeerreviewmaketitle

\IEEEraisesectionheading{\section{Introduction}\label{sec:introduction}}

\IEEEPARstart{C}{ontainers} have been widely adopted on Cloud systems. Applications together with their dependencies are encapsulated into containers \cite{8360359}, which can ensure environment compatibility and enable users to move and deploy programs easily among clusters. Containerisation is a virtualisation technology \cite{DBLP:journals/spe/RodriguezB19}. Rather than creating an entire operating system (called guest OS) on top of a host OS as in a Virtual Machine (VM), containers only share the host kernel, which makes containers more lightweight than VMs. Containers on Cloud are often dedicated to run \textit{micro-services} \cite{8457916} and one container mostly hosts one application or a part of it. 

High Performance Computing (HPC) systems are traditionally employed to perform large-scale financial, engineering and scientific simulations \cite{OlivierTerzo2022} that demand low latency (\textit{e.g.} interconnect) and high throughput (\textit{e.g.} the number of jobs completed over a specific time). To satisfy different user requirements, HPC systems normally provide predefined modules with specific software versions that users can switch by loading or unloading the modules with the desired packages \cite{6114480}. This approach requires assistance of system administrators and therefore limits increasing user demands for environment customisation. On a multi-tenant environment as on HPC systems, especially HPC production systems, installation of new software packages on-demand by users is restricted, as it may alter the working environments of existing users and even raise security risks. Module-enabled software environments are also inconvenient for dynamic Artificial Intelligence (AI) software stacks \cite{8916576}. Big Data Analytics hosted on Cloud are compute-intensive or data-intensive, mainly due to deployments of AI or Machine Learning (ML) applications, which demand extremely fast knowledge extraction in order to make rapid and accurate decisions. HPC-enabled AI can offer optimisation of supply chains, complex logics, manufacturing, simulation and underpin modelling to solve complex problems \cite{Yi2019}. Typically, AI applications have sophisticated requirements of software stacks and configurations. Containerisation not only enables customised environments on HPC systems, but also brings research reproducibility into practice.

Containerised applications can become complex, \textit{e.g.} thousands of separate containers may be required in production, and containers may require network isolation among each other for security reasons. Sophisticated strategies for container orchestration  \cite{10.4108/eai.25-10-2016.2266649} have been developed on Cloud or big-data clusters to meet such requirements. HPC systems, \textit{per contra}, lack features of efficiency in container scheduling and management (\textit{e.g}. load balancing and auto container scaling), and often provide no integrated support for environment provisioning (\textit{i.e.} infrastructure, configurations and dependencies).

There have been numerous studies on containerisation and container orchestration on Cloud \cite{10.1109/CISIS.2015.35, DBLP:journals/spe/RodriguezB19, Pahl2017, Casalicchio2019, 7868429, Bernstein2014, 10.1145/3378447, doi:10.1002/cpe.5668}, however, there is no comprehensive survey on these technologies and techniques for HPC systems existing as of yet. This article:
\begin{itemize}
	\item Investigates state-of-the-art works in containerisation on HPC systems and underscores their differences with respect to the Cloud;
	\item Introduces the representative orchestration frameworks on both HPC and Cloud environments, and highlights their feature differences; 
	\item Gathers the related studies in the integration of container orchestration strategies on Cloud into HPC environments; 
	\item Discusses the challenges and envisions the potential directions for research and engineering.  
\end{itemize}

The rest of the paper is organised as follows. First, Section~\ref{sec:containerisation_tech} introduces the background on containerisation technologies and techniques. Key technologies of state-of-the-art container engines (Section~\ref{subsec:containers}) and orchestration strategies (Section~\ref{sec:orchestration}) are presented, and the feature differences thereof between HPC and Cloud systems are discussed. Next, Section~\ref{sec:vision_challenge} describes research challenges and the vision. Lastly, Section~\ref{sec:conclude} concludes this paper.

\section{Concepts and Technologies for Containerisation}\label{sec:containerisation_tech}

The main differences between containerisation technologies on Cloud and HPC systems are in terms of security and the types of workloads. The HPC applications tend to require more resources as to not only CPUs, but also the amount of memory and network speed. HPC communities have, therefore, developed sophisticated workload managers to leverage hardware resources and optimise application scheduling. Since the typical applications on Cloud differ significantly from those in HPC centres with respect to the sizes, execution time and requirements of the availability of hardware resources \cite{10.1145/3150224}, the management systems on Cloud are evolved to include architectures different from those on HPC systems. 

Research and engineering on containerisation technologies and techniques for HPC systems can be classified into two broad categories: 
\begin{enumerate}
	\item Container engines/runtimes;
	\item Container orchestration. 
\end{enumerate}
In the first category, various architectures of container engines have been developed which vary in usage of namespaces (see Section~\ref{subsec:container_concepts}), image formats and programming languages. The research in the latter category is still in its primitive stage, which will be discussed in Section~\ref{sec:orchestration}.
\subsection{Containerisation Concepts}\label{subsec:container_concepts}
\begin{figure}[!t]
	\centering
	\includegraphics[width=0.45\textwidth]{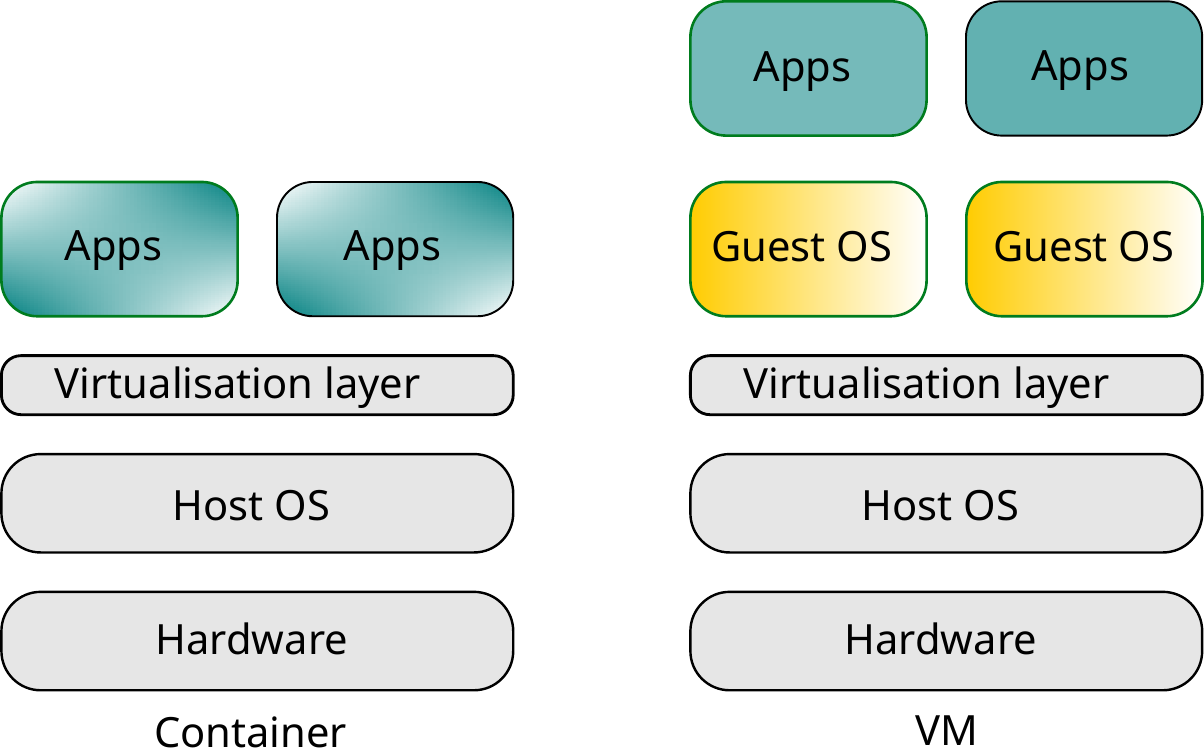}
	\caption[]{Structure comparison of VMs and containers. On the VM side, the virtualisation layer often appears to be hypervisor while on the container side it is the container runtimes.}
	\label{fig:structure_virtualization}
\end{figure}

Containerisation is an OS-level virtualisation technology \cite{Merkel2014} that provides separation of application execution environments. A container is a runnable instance of an image that encapsulates a program together with its libraries, data, configuration files, \textit{etc.} \cite{8360359} in an isolated environment, hence it can ensure library compatibility and enables users to move and deploy programs easily among clusters. A container utilises the dependencies in its host kernel. The host merely needs to start a new process that is isolated from the host itself to boot a new container \cite{7933304}, thus making container start-up time comparable to that of a native application. In contrary, a traditional VM loads an entire guest kernel (simulated OS) into memory, which can occupy gigabytes of storage space on the host and requires a significant fraction of system resources to run. VMs are managed by \textit{hypervisor} which is also known as Virtual Machine Monitor (VMM) that partitions and provisions VMs with hardware resources (\textit{e.g.} CPU and memory). The hypervisor gives the hardware-level virtualisation \cite{10.1145/2988336.2988337, 7092949}. Fig.~\ref{fig:structure_virtualization} highlights the architecture distinction of VMs and containers. It is worth noting that containers can also run inside VMs \cite{vmware2018}. Besides portability, containers also enable reproducibility, \textit{i.e}. once a program has been defined inside the container, its included working environment remains unchanged regardless of its running occurrences. Nevertheless, the shared kernel strategy presents an obvious pitfall: a Windows containerised application cannot execute on Unix kernels. Obviously, this should not become an impediment to its usage as Unix-like OS are often the preference for HPC systems.

HPC applications are often highly optimised for processor architectures, interconnects, accelerators and other hardware aspects. Containerised applications, therefore, need to compromise between performance and portability. The studies have shown that containers can often achieve near-native performance \cite{7933304,8820966, Hu2019, Younge2017, 10.1145/3147213.3147231, 10.1186/s13673-017-0124-3, 9284294} (see Section~\ref{subsubsec:performance}).

Linux has several namespaces \cite{10.1145/3126908.3126925} that isolate various kernel resources: \texttt{mount} (file system tree and mounts), \texttt{PID} (process ID), \texttt{UTS} (hostname and domain name), \texttt{network} (\textit{e.g.} network devices, ports, routing tables and firewall rules), \texttt{IPC} (inter-process communication resources) and \texttt{user}. The last namespace is an unprivileged namespace that grants the unprivileged process access to traditional privileged functionalities under a safe context. More specifically, the \texttt{user} namespace allows to map user ID (UID) and group ID (GID) from hosts to containers, meaning that a user having UID 0 (root) inside a container can be mapped to a non-root ID (\textit{e.g.} 100000) outside the container. \texttt{Cgroups} (Control Groups) is another namespace that is targeted to limit, isolate and measure resource consumption of processes. \texttt{Cgroups} is useful for a multi-tenant setting as excess resource consumption of certain users will be only adverse to themselves. One application of Linux namespaces is the implementation of containers, \textit{e.g.} Docker, the most widely-used container engine, uses namespaces to provide the isolated workspace that is called \textit{container}. When a container executes, Docker creates a set of namespaces for that container.

\subsection{Docker}\label{subsubsec:docker}

There are multiple techniques that realise the concept of containers. Docker is among the most popular ones \cite{9284294}. After its appearance in 2013, various container solutions aimed for HPC have emerged \cite{8820966}. Docker, initially based on LXC \cite{SenthilKumaran2017}, is a container engine that supports multiple platforms, \textit{i.e.} Linux, OSX and Windows. A Docker container image is composed of a readable/writable layer above a series of read-only layers. A new writable layer is added to the underlying layers when a new Docker container is created. All changes that are made to the running container, such as writing new files, modifying or deleting existing files, are written to this thin writable container layer. Docker adopts namespaces including \texttt{Cgroups} to provide resource isolation and resource limitation, respectively. Table~\ref{table:namespace_list} highlights the usage of namespaces with respect to Docker and a list of container engines targeted for HPC environments.

\begin{table*}[!ht]
	\caption[]{Linux namespace supports for HPC-targeted container engines (Section~\ref{subsec:containers}) and Docker in the year of 2022. \\Note that without certain namespaces, containers may still operate however with restricted functionalities.}\label{table:namespace_list}
	\centering  
	\tabulinesep=0.8mm	
	\begin{tabu}to \textwidth {X|X|X|X|X|X|X}   
	  \toprule
		\textbf{Namespaces} & \textbf{Singularity}& \textbf{Shifter}&\textbf{Charliecloud}&\textbf{UDocker} &\textbf{ SARUS}&\textbf{Docker}\\
		
		\toprule
		\texttt{mount} &\cmark  &\cmark & \cmark& \cmark &\cmark &\cmark \\
		\hline
		\texttt{PID}  & \cmark&\xmark & \xmark &\xmark &\xmark&\cmark\\
		\hline
		\texttt{UTS}  &  \xmark&\xmark & \xmark&\xmark&\xmark&\cmark\\
		\hline
		\texttt{network } &\xmark  & \xmark& \xmark&\xmark&\xmark&\cmark\\
		\hline
		\texttt{IPC}  &\xmark  &\xmark & \xmark&\xmark&\xmark&\cmark\\
		\hline
		\texttt{user} & \cmark & \cmark& \cmark&\cmark&\cmark &\cmark\\
		\hline
		\texttt{Cgroup}  & \cmark & \cmark& \xmark&\xmark&\cmark&\cmark\\[1ex]		
		
		\bottomrule	
			\end{tabu}
		\\[1ex] 
\end{table*}

Docker provides network isolation and communication by creating three types of networks: \textit{host}, \textit{bridge} and \textit{none}. The bridge network is the default Docker network. The Docker engine creates a subset or gateway to the bridged network. This software \textit{bridge} allows Docker containers to communicate within the same bridged network; meanwhile, isolates the containers from a different bridged network. Containers in the same host can communicate via the default network by the host IP address. To communicate with the containers located on a different host, the host needs to allocate ports on its IP address. Managing ports brings overhead which can intensify at scale. Dynamically managing ports can solve this issue which is better handled by orchestration platforms as introduced in Section~\ref{subsec:orch_cloud}.

Docker is widely adopted in Cloud where users often have root privileges. The root privilege is required to execute the Docker application and its Daemon process that provides the essential services. Originally running Docker with root permission brings some advantages to Cloud users. For instance, users can run their applications and alternative security modules to provide separation among different allocations \cite{Saha2018}; users can also mount host filesystems to their containers. Root privilege can cause security issues. Therefore, the latest updates of Docker engine start to support rootless daemon and enable users to execute containers without root. Nevertheless, other security concerns still persist. For instance, usage of Unix socket can be changed to TCP socket which will grant an attacker a remote control to execute any containers in the privileged mode. Additionally, rootless Docker does not run out of box, system administrators need to carefully set the namespaces of hosts to separate resources and user groups in order to guarantee security. Hence HPC centres that typically have high security requirements are still reluctant to enable the Docker support on their systems. 

\section{Container Engines and Runtimes for HPC Systems}\label{subsec:containers}
This section first reviews the state-of-the-art container engines/runtimes designed for HPC systems and compares the major differences with the mainstream Cloud container engine, \textit{i.e.} Docker. Next, Section~\ref{subsubsec:performance} shows the performance evaluation of the reviewed HPC container engines.

\subsection{State-of-the-Art Container Engines and Runtimes}	
A list of representative container engines and runtimes for HPC systems is given in this section. They differ in functional extent and implementation, however, also hold some similarities. Table~\ref{table:namespace_list} and Table~\ref{table:other_feature_list} summarise the feature differences and similarities between Docker and a list of main HPC container engines.

\subsubsection{Shifter}\label{subsubsec:shifter}
Shifter \cite{Gerhardt_2017} is a prototypical implementation of container engine for HPC developed by NERSC. It utilises Docker for image building workflow. Once an image is built, users can submit it to an unprivileged gateway which injects configurations and binaries, flattens it to an ext4 file system image, and then compresses to squashfs images that are copied to a parallel filesystem on the nodes. In this way, Shifter insulates the network filesystem from image metadata traffic. Root permission of Docker is naturally deprived from Shifter that only grant user-level privileges. Existing directories can be also mounted inside Shifter image by passing certain flags.

As an HPC container engine, Shifter supports MPICH that is an implementation of the Message Passing Interface (MPI) \cite{Gropp:1994:UMP:207387,mpistandardv3.1} standard. To enable accelerator supports such as GPU without compromising container portability, Shifter runtime swaps the built-in GPU driver of a Shifter container with an ABI (Application Binary Interface) compatible version at the container start-up time.

\begin{table*}[!ht]
\caption[]{Comparison of Docker with the list of container engines for HPC systems. \\WLM: workload manager. Orchestration is described in Section~\ref{subsection:wlm_hpc} and Section~\ref{subsec:orch_cloud}.}\label{table:other_feature_list}. 
\centering  
\tabulinesep=0.8mm		
\begin{tabu}to \textwidth {l|X| X| X| X| X |l}   

	\toprule 	
\textbf{Container engines} &\textbf{Docker} & \textbf{Singularity}&  \textbf{Shifter }& \textbf{Charliecloud} & \textbf{SARUS} &\textbf{UDocker}\\
	\toprule
	Usage of namespaces	& \cmark& \cmark& \cmark& \cmark& \cmark &\cmark\\
	\hline 
	
	MPI support	& \cmark& \cmark& \cmark& \cmark& \cmark &\cmark\\
	\hline
	GPU support & \cmark& \cmark& \cmark& \cmark& \cmark &\cmark\\
	\hline
	Network support &Pluggable network driver (\textit{e.g.} bridge) &Host network & Host network &Host network& Host network &Host network	\\
	\hline
	Image format & Layers of files &Single image file, Filesystem bundle&squashfs layers &Filesystem bundle&Filesystem bundle&layers of files\\
	\hline
	Access to host filesystems  &\cmark &\cmark & \cmark&\cmark & \cmark&\cmark\\
	\hline
	Escalation of permission  &\cmark&\xmark&\xmark&\xmark&\xmark&\xmark\\
	\hline 
	Privileged daemon* &\cmark&\xmark&\xmark&\xmark&\xmark&\xmark\\
	\hline
	Orchestration & Docker Swarm &HPC WLM &HPC WLM &HPC WLM&HPC WLM& HPC WLM\\
	\hline
	Programming languages & Go& Go& C	&C&C++&Python\\
	\toprule
	
\end{tabu}
\\[1ex] 
*Starting from v19.03, Docker also provides options to change its daemon to be rootless.
\end{table*}

\subsubsection{Charliecloud}
Charliecloud \cite{10.1145/3126908.3126925} runs containers without privileged operations or daemons. Charlicloud can convert a Docker image into a tar file and unpacks it on the HPC nodes. Installation of Charliecloud does not require root permission. Such non-intrusive mechanisms are ideal for HPC systems. Charliecloud is considered to be secure against \textit{shenanigans}, such as \texttt{chroot} escape, bypass of file and directory permission, privileged ports bound to all host IP addresses or UID set to an unmapped UID \cite{doi:10.1002/cpe.5668}. 

MPI is supported by Charliecloud. Injecting host files into images is used by Charliecloud to solve library compatibility issues, such as GPU libraries that may be tied to specific kernel versions.     

\subsubsection{Singularity}

Singularity is the most-widely used HPC container engine in academia and industry. Singularity \cite{Kurtzer2017SingularitySC} was specifically designed from the outset for HPC systems. Contrasting with Docker, it gives the following merits \cite{Hu2019}:
\begin{itemize}
\item Running with user privileges and no daemon process. Only user privileges are required to execute Singularity applications. Acquisition of root permission is only necessary when users want to build or rebuild images, which can be performed on their own working computers. Unprivileged users can also build an image from a definition file with a few restrictions by "fake root" in Singularity, however, some methods requiring to create block devices (\textit{e.g.} \texttt{/dev/null}) may not always work correctly in this way;
\item Seamless integration with HPC systems. Singularity natively supports GPU, MPI and InfiniBand \cite{10.1145/3150224}. No additional network configurations are expected in contrast with Docker containers; 
\item Portable via a single image file (SIF format). On the contrary, Docker is built up on top of layers of files.
\end{itemize}

Two approaches are often used to execute MPI applications using Singularity, \textit{i.e.} hybrid model and bind model. The former compiles MPI binaries, libraries and the MPI application into a Singularity image. The latter binds the container on a host location where the container utilises the MPI libraries and binaries on the host. The latter model has a smaller image size since it does not include compiled MPI libraries and binaries in the image. Utilising the host libraries is also beneficial to application performance, however, the version of MPI implementation that is used to compile the application inside the container must be compatible with the version available on the host. The hybrid model is recommended, as mounting storage volumes on the host often require privileged operations. 

Most Docker images can be converted to singularity images directly via simple command lines (\textit{e.g}. \texttt{docker save}, \texttt{singularity build}). Singularity has quickly become the \textit{ipso facto} standard container engine for HPC systems.

\subsubsection{SARUS}
\begin{figure}[!t]
\centering
\includegraphics[width=0.5\textwidth]{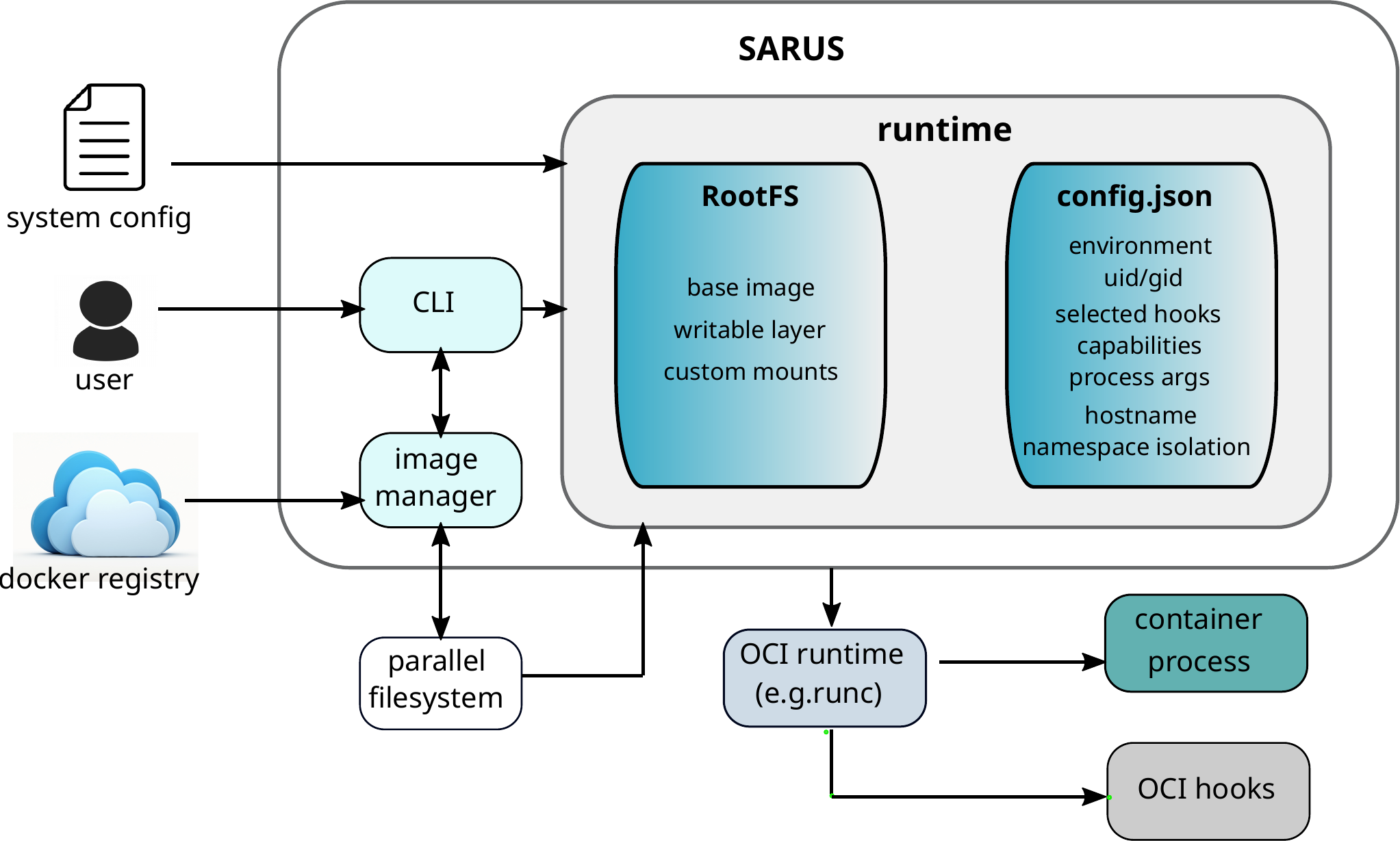}
\caption[]{The internal structure of SARUS. OCI hooks include MPI hook.}
\label{fig:sarus_arch}
\end{figure}

SARUS \cite{10.1007/978-3-030-34356-9_5} is another container engine targeted for HPC systems. SARUS relies on \texttt{runc}\footnote{ \url {https://github.com/opencontainers/runc}} to instantiate containers. \texttt{runc} is a CLI (Command-Line Interface) tool for spawning and running containers according to the OCI (Open Container Initiative) specification. Different from the aforementioned engines, the internal structure of SARUS is based on the OCI standard (see Section~\ref{subsubsec:issue_runtime_image}). As shown in Fig.~\ref{fig:sarus_arch}, the CLI component takes the command lines which either invoke the \textit{image manager} component or the \textit{runtime} component. The latter instantiates and executes containers by creating a bundle that comprises a root filesystem directory and a JSON configuration file. The \textit{runtime} component then calls \texttt{runc} that will spawn the container processes. It is worth noting that functionalities of SARUS can be extended by calling customised OCI hooks, \textit{e.g.} MPI hook.

\subsubsection{UDocker}\label{subsubsec:udocker}
UDocker\footnote{\url{https://github.com/indigo-dc/udocker}} is a Python wrapper for the Docker container, which executes only simple Docker containers in user space without the acquisition of root privileges. UDocker provides a Docker-like CLI and only supports a subset of Docker commands, \textit{i.e.} \texttt{search}, \texttt{pull}, \texttt{import}, \texttt{export}, \texttt{load}, \texttt{save}, \texttt{create} and \texttt{run}. It is worth noting that UDocker neither makes use of Docker nor requires its presence on the host. It executes containers by simply providing a chroot-like environment over the extracted container.

\subsubsection{Other HPC Container Engines}\label{subsubsec:other_HPC_container}	
More and more HPC container engines are being developed, this section gives an overview of some that are targeted for special use cases.

Podman \cite{10.1007/978-3-030-59851-8_23} makes use of the \texttt{user} namespace to execute containers without privilege escalation. A Podman container image comprises layers of read-write files as Docker. It adopts the same runtime \texttt{runc} as in SARUS and Docker. The runtime \texttt{crun}, which is faster than \texttt{runc}, is also supported. A notable feature of Podman is as its name denotes: the concept of \textit{pod}. A pod groups a set of containers that collectively implements a complex application to share namespaces and simplify communication. This feature enables the convergence with the Kubernetes \cite{10.5555/3175917} environment (Section~\ref{subsec:kubernetes}), however, requires advanced kernel features (\textit{e.g.} version 2 \texttt{Cgroups} and user-space FUSE). These kernel features are not yet compatible with network filesystems to make full use of the rootless capabilities of Podman and consequently restrains its usage from HPC production systems \cite{ruhela-TexasScholarWorks} .

Similar to UDocker, Socker \cite{7923813} is a simple secure wrapper to run Docker in HPC environments, more specifically SLURM (Section~\ref{subsec:slurm}). It does not support the \texttt{user} namespace, however, it takes the resource limits imposed by SLURM.  

Enroot\footnote{\url{https://github.com/NVIDIA/enroot}} from NVIDIA can be considered as an enhanced unprivileged \texttt{chroot}. It removes much of the isolation that the other container engines normally provide but preserves filesystem separation. Enroot makes use of \texttt{user} and \texttt{mount} namespaces.

\subsection{Performance Evaluation for HPC Container Engines}\label{subsubsec:performance}

This section only selects the representative works as given in Table~\ref{table:container_performance_hpc}, rather than exhausting the literature, to show the performance of containers that are specifically targeted for HPC systems in terms of CPU, memory, disk (I/O), network and GPU. Table~\ref{table:benchmark_hpc} lists the benchmarks utilised in these work. Overall, the container startup latency can be high on the Cloud. This startup overhead is caused by building containers from multiple image layers, setting read-write layers and monitoring containers \cite{9284294}. An HPC container is composed of a single image or directory (with exception to Podman) and monitoring is performed by HPC systems.  

\begin{table*}[ht]
	\caption[]{Overview of the related work on container performance evaluation in terms of CPU, memory, disk, network and GPU on HPC systems.}\label{table:container_performance_hpc}
	\centering  
		\tabulinesep=0.8mm
		\begin{tabu}to \textwidth{ l| X| X|  X |X} 
			\toprule	
			
			\textbf{Metrics}	&\textbf{Performance Overhead }&\textbf{Work}&\textbf{ Container engines} & \textbf{HPC vendors}\\
			\bottomrule
			CPU time & Often negligible. Large overhead caused by vendor-tuned libraries and dynamically linking libraries; better performance in many-process Python programs& \cite{Younge2017, Hu2019, 7933304, 8820966, 9284294, Bahls2016}& Singularity, Shifter, Charliecloud, Podman, SARUS  &Cray XC, Cray XE/XK  \\
		
			\hline 
			Memory usage &Negligible&\cite{Hu2019, 9284294}  &Singularity, Charliecloud, Podman &-\\
			\hline
			Disk usage  &Negligible& \cite{9284294} &  Singularity,  Charliecloud, Podman &-\\ 
			\hline
			Network &Negligible or slight overhead. Overhead happens at start-up time because of the single file/bundle structure&\cite{Younge2017, Hu2019, 9284294} &  Singularity, Charliecloud, Podman  &Cray XC \\
			\hline 
			GPU &A slight overhead&\cite{Hu2019, 10.1007/978-3-030-34356-9_6} 	&Singularity    &IBM\\ 
			\bottomrule			
				
		\end{tabu}	
		\\[1ex]
		
	\end{table*}

\begin{table}[ht]
	\caption[]{The list of HPC benchmarks mentioned in Section~\ref{subsubsec:performance}.}\label{table:benchmark_hpc}
	\centering  
	
			\tabulinesep=0.8mm
		\begin{tabu}to 0.5\textwidth{ l| X} 
			\toprule	
			
			\textbf{Benchmarks}	&\textbf{Description}\\
			\bottomrule
			IMB& Intel MPI Benchmark   \\
			
			HPCG & A complement to the LINPACK  \\ 	
	
			Linpack  &Measure floating-point computing power \\ 
	
		NAMD & Simulation for molecular dynamics\\
		
			VASP &Atomic scale materials modelling\\ 
				
			WRF & Weather Research and Forecasting Model \\
			AMBER & Assisted Model Building with Energy Refinement\\
			HPGMG-FE & High-Performance Geometric Multigrid, Finite Element\\
			\bottomrule	
	\end{tabu}

	\end{table}

The work in \cite{Younge2017}, utilising the IMB \cite{IMBbenchmark} benchmark suite and HPCG \cite{Dongarra2015} benchmarks, proved that little overhead of network bandwidth and CPU computing overhead is caused by Singularity when dynamically linking vendor MPI libraries in order to efficiently leverage advanced hardware resources. With the Cray MPI library, Singularity container achieved 99.4\% efficiency of native bandwidth on a Cray XC \cite{crayxc40} HPC testbed when running the IMB benchmark suite. However, the efficacy drastically drops to 39.5\% with Intel MPI. Execution time evaluated with the HPCG benchmarks, indicated that the performance penalty caused by Singularity is negligible with Cray MPI, though the overhead can reach 18.1\% with Intel MPI. The performance degradation with Intel MPI is mostly because of the vendor-tuned MPI library which does not leverage hardware resources from a different vendor, \textit{e.g.} interconnect.

Hu \textit{et al.} \cite{Hu2019} evaluated the Singularity performance in terms of CPU capacity, memory, network bandwidth and GPU with Linpack benchmarks \cite{doi:10.1137/1.9781611971811} and four typical HPC applications (\textit{i.e.} NAMD \cite{NAMDbenchmark}, VASP \cite{VASPbenchmark}, WRF \cite{WRFbechmark} and AMBER \cite{AMBERbenchmark}). Singularity provides close to native performance on CPU, memory and network bandwidth. A slight overhead (4.135\%) is shown on NVIDIA GPU. 

Muscianisi \textit{et al.} \cite{10.1007/978-3-030-34356-9_6} illustrated the performance impact of Singularity with the increasing number of GPU nodes. The evaluation was carried out on CINECA's GALILEO systems with TensorFlow \cite{Abadi2016} applications. The results again demonstrated that the container environments caused negligible performance overhead.

The work by Hale \textit{et al.} \cite{7933304} presented the CPU performance of Shifter with HPGMG-FE (MPI implementation) benchmarks \cite{HPGMGbenchmark} on Cray XC30 (192 cores, 24 cores per compute node) where the performance margin between Shifter container and bare metal is unnoticeable. Comparison is also given for MPI with implementation in C++ and Python using a custom benchmark. The authors observed that it could take over 30 minutes to import the Python modules when running natively with 1,000 processes. Each process of a Python application imports modules from the filesystem on each node. Accesses to many small files on an HPC filesystem using many processes can be extremely slow comparing with the accesses to a few large files. The containerised benchmark has already included all the modules in its image that is mounted as a single file on each node, therefore, Shifter container outperforms the native execution in this case. \textit{Bahls} \cite{Bahls2016} also evaluated the execution time of Shifter on Cray XC and Cray XE/XK systems exploiting Cray HSN (High Performance Network). Their results showed that Shifter gave comparable performance to bare metal. 

The study in \cite{8820966} compared the performance of Shifter and Singularity against bare metal in terms of computation time using two biological use cases on three types of supercomputer CPU architectures: Intel Skylake, IBM Power9 and Arm-v8. Containerised applications can scale at the same rate as the bare-metal counterparts. However, the authors also observed that with a small number of MPI ranks, containers should be built as generic as possible, \textit{per contra}, when it comes to a large number of cores, containers need to be tuned for the hosts. 

Without performance comparison with bare-metal applications, the work in \cite{9284294} studied the CPU, memory, network and I/O performance of Charliecloud, Podman and Singularity. All the containers behave similarly with respect to the CPU and memory usage. Charliecloud and Singularity have comparable I/O performance. Charliecloud incurs large overhead on Lustre's MDS (Metadata Server) and OSS (Object Storage Server) due to its bare tree structure. Comparing with the structures of shared layers (as in Docker), this structure needs to access a large number of individual files from the image tree from Lustre. Consequently, it causes network overhead when data is transmitted from the client node over the network at container start-up time. Similarly, as Singularity is stored as a single file on Lustre, a large amount of data needs to be loaded at starting point resulting in a data transmission spike on network. 	 

SARUS has shown strong scaling capability on Cray XC systems with hybrid GPU and CPU nodes \cite{10.1007/978-3-030-34356-9_5}. The performance difference between SARUS and bare metal is less than 0.5\% up to 8 nodes and 6.2\% up to 256 nodes. No specific metrics are given in terms of GPU, though GPU has been used as accelerators.

\subsection{Section Highlights}
Containers are introduced to HPC systems, as they enable environment customisation for users, which offers the solutions to application compatibility issues. This is particularly important on HPC systems that are typically inflexible for environment modifications. Notably, HPC container engines are designed to meet the high-security requirements on HPC systems. Multiple prevailing engines have been described in this section, they share some common features: 
\begin{itemize}
	\item Non-root privileges;	
	\item Often can convert Docker images to their own image formats;
	\item Supports of MPI that are typical HPC applications;
	\item Use host network rather than pluggable network drivers.		 	 
\end{itemize}
Yet differences exist in their image formats. Layered image format is seen in Docker (UDocker wraps Docker image layers to a local directory), which is executed by pulling the image layers that have not been previously downloaded on the host. HPC container images are stored in a single directory or file which can be transferred to the compute nodes easily avoiding the pulling operations that require network access. HPC container engines show various ways to incorporate well-tuned libraries targeting for the hosts in order to achieve optimised performance, \textit{e.g.} OCI hooks (SARUS), injecting host files into images (Charliecloud). 

Section~\ref{subsubsec:performance} aims to give examples that can provide general advices on how to build the container images to maximise performance. Clearly, performance loss can occur in certain cases which are summarised in the second column of Table~\ref{table:container_performance_hpc}.

\section{Container Orchestration}\label{sec:orchestration}	
\textit{Orchestration} under the context herein means automated configuration, coordination and management of Cloud or HPC systems. In theory, HPC workload manager can be also addressed as orchestrator, however, this article takes the former term as it is the custom terminology that has been long-used and widely understood in the HPC area. The driving factors that push HPC workload managers and Cloud orchestrators to be developed in different directions can be multiple. This will be discussed at the end of this section (Section~\ref{subsec:highlights}). However, first it is important to understand the mechanisms of HPC workload managers (Section~\ref{subsection:wlm_hpc}) and Cloud orchestrators (Section~\ref{subsec:orch_cloud}). Mostly, container orchestration for HPC systems either relies on the orchestration strategies of the existing Cloud orchestrators or exploits the mechanisms of current HPC workload managers or software tools. This point will be depicted in Section ~\ref{sec:stateart}.

\subsection{Workload Managers for HPC Systems}\label{subsection:wlm_hpc}
Cloud aims to exploit economy of scale by consolidating applications into the same hardware \cite{10.1145/3150224} and the hardware resources can be easily extended based on user demands. In contrast, HPC centres have large-scale hardware resources available and reserve computing resources exclusively for users. Table~\ref{table:comp_hpc_cloud_orchestrator} underscores the main differences between HPC workload managers and Cloud orchestrators. A typical HPC system is managed by a workload manager. A \textit{workload manager} comprises a \textit{resource manager} and a \textit{job scheduler}. A resource manager \cite{Hovestadt2003} allocates resources (\textit{e.g.} CPU and memory), schedules jobs and guarantees no interference from other user processes. A job scheduler determines the job priorities, enforces resource limits and dispatches jobs to available nodes \cite{Klusacek2015}. 

HPC workload managers incorporate a big family, such as PBS \cite{Staples2006}, Spectrum LSF \cite{10.1002/spe.4380231203}, Grid Engine \cite{923173}, OAR \cite{10.5555/1169223.1169583} and Slurm \cite{Jette02slurm:simple}. Slurm and PBS are two main-stream workload managers. The workload managers shares some common features: a centralised scheduling system, a queuing system and static resource management mechanisms, which will be detailed in this section.

\begin{table*}[ht]
\caption[]{Comparison of HPC workload managers (Section~\ref{subsection:wlm_hpc}) and cloud orchestrators (Section~\ref{subsec:orch_cloud}).}\label{table:comp_hpc_cloud_orchestrator}
\centering  

	\tabulinesep = 0.8mm
		\begin{tabu}to \textwidth{@{} l| X |X@{} }				
				\toprule		
				& \textbf{HPC workload manager}& \textbf{Cloud orchestrator}\\
				\toprule		
				Deployment &Batch queue (queueing time from seconds to days) & Often immediate \\
				\hline
				Workload type& Binary& Container, pod \\	
				\hline
				Supports of Parallel and Array Jobs  & Both & Array*  \\
				\hline
				Resource unit & Bare-metal nodes & Pods, VM nodes\\ 
				\hline 
				Resource elasticity & No& Yes\\
				\hline	
				Application execution length &Long duration \& Run to completion & Continuously running or short duration$^{\dagger}$ \\ 	
				\hline				
				Application specifics & Distributed memory jobs (\textit{e.g.} MPI) & Often micro-services\\	
				
				\hline
				DevOps environment provision& No& Yes \\
				\hline
				API supports & No (or very weak)& Yes\\
				\hline 
				Job scheduling & Backfilling& On-demand scheduling\\
				\hline 
				Centralised scheduling system & Yes & Not always \\
				\hline 
				Job submission scripts &batch scripts & Declarative files, typically \texttt{yaml} scripts \\
				\hline
				Checkpointing & Yes & No. Containers are relaunched upon failure\\
				\hline
				Support of multiple resource managers  & No& Often yes\\		
				\bottomrule		
			\end{tabu}	
			\\[1ex] 
		Exceptions: *Mesos can support parallel jobs (Section~\ref{subsubsec:mesos}); $^{\dagger}$YARN targets for long-running batch jobs (Section~\ref{subsubsec:mesos})	
		\end{table*}

		\subsubsection{PBS}\label{subsec:torque}
		
		\begin{figure}[!t]
			\centering
			\includegraphics[width=0.45\textwidth]{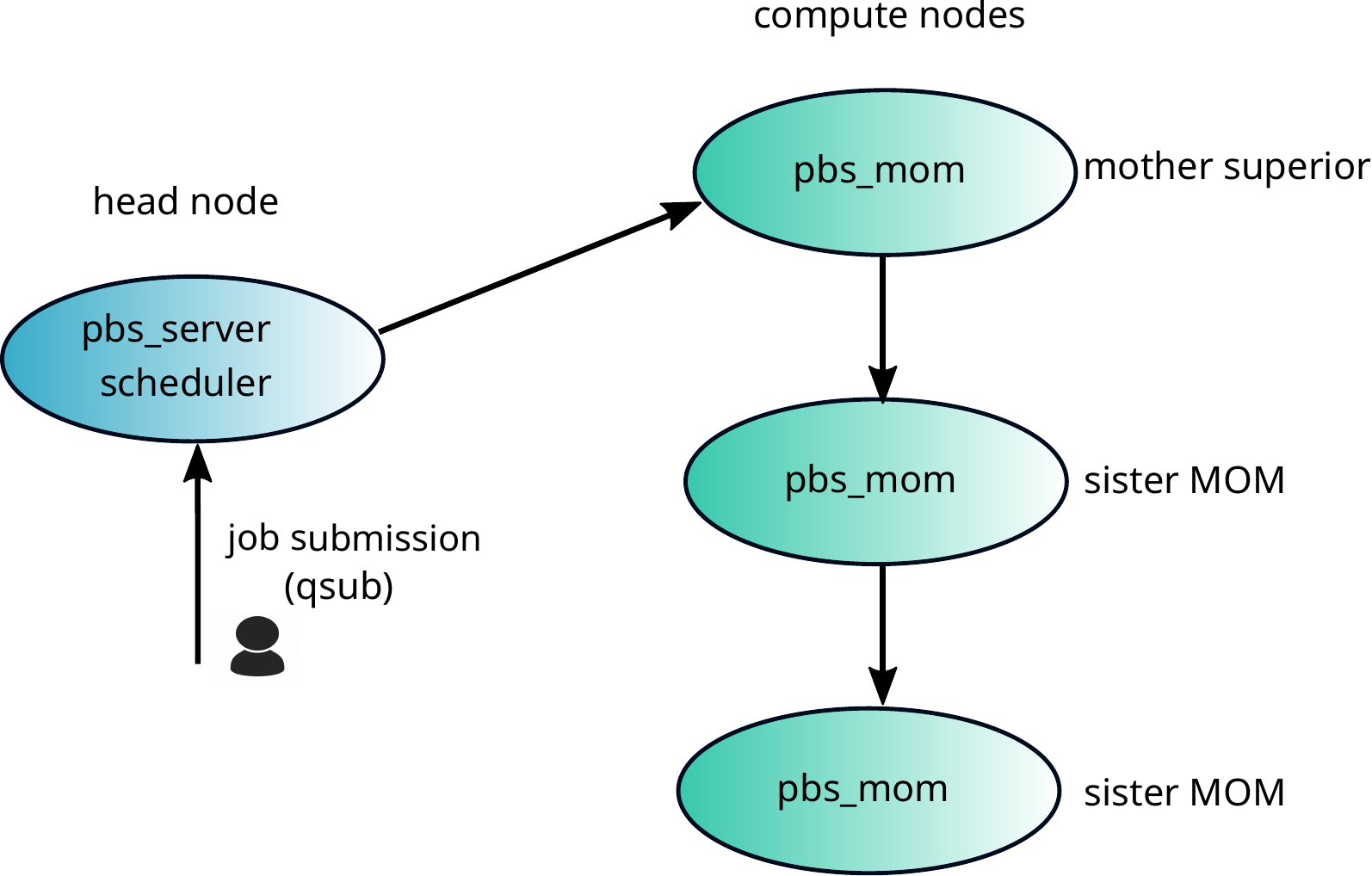}
			
			\caption[]{TORQUE structure. $pbs\_server$, scheduler and $pbs\_mom$ are the daemons running on the nodes. Mother Superior is the first node on the node list (on step4).}
			\label{fig:torque_structure}
		\end{figure}

		PBS stands for Portable Batch System which includes three versions: OpenPBS, PBS Pro and TORQUE. OpenPBS is open-source and TORQUE is a fork of OpenPBS. PBS Pro is dual-licensed under an open-source and commercial license. The structure of a TORQUE-managed cluster consists of a head node and many compute nodes as illustrated in Fig.~\ref{fig:torque_structure} where only three compute nodes are shown. The head node (coloured in blue in Fig.~\ref{fig:torque_structure}) controls the entire TORQUE system. A $pbs\_server$ daemon and a job scheduler daemon are located on the head node. The batch job is submitted to the head node (in some cases, the job is first submitted to a login node and then transferred to the head node). A node list that records the configured compute nodes is maintained on the head node. The architecture of this kind as shown in Fig~\ref{fig:torque_structure} represents the fundamental cluster structure of main-stream HPC workload managers. The procedure of job submission on TORQUE is briefly described as follows:
		\begin{enumerate}
			\item The job is submitted to the head node by the command \texttt{qsub}. A job is normally written in the format of a PBS script. A job ID is returned to the user as the standard output of \texttt{qsub}.  
			
			\item The job record, which incorporates a job ID and the job attributes, is generated and passed to $pbs\_server$.
			\item $pbs\_server$ transfers the job record to the job scheduler daemon. The job scheduler daemon adds the job into a job queue and applies a scheduling algorithm to it (\textit{e.g.} FIFO: First In First Out) which determines the job priority and its resource assignment.
			
			\item When the scheduler finds the list of nodes for the job, it returns the job information to $pbs\_server$. The first node on this list becomes the \textit{Mother superior} and the rest are called \textit{sister MOMs} or \textit{sister nodes}. $pbs\_server$ allocates the resources and passes the job control as well as execution information to the $pbs\_mom$ daemon installed on the mom superior node instructing to launch the job on the assigned compute nodes. 
			\item The $pbs\_mom$ daemons on the compute nodes manage the execution of jobs and monitor resource usage. $pbs\_mom$ will capture all the outputs and direct them to \textit{stdout} and \textit{stderr} which are written into the output and error files and are copied to the designated location when the job completes successfully. The job status (completed or terminated) will be passed to $pbs\_server$ by $pbs\_mom$. The job information will be updated.
		\end{enumerate}

		In TORQUE, nodes are partitioned into different groups called $queues$. In each queue, the administrator sets limits for resources such as walltime and job size. This feature can be useful for job scheduling in a large HPC cluster where nodes are heterogeneous or certain nodes are reserved for special users. This feature is commonly seen in HPC workload managers.
		
		TORQUE has a default scheduler FIFO, and is often integrated with a more sophisticated job scheduler, such as Maui \cite{10.1007/3-540-45540-X_6}. Maui is an open source job scheduler that provides advanced features such as dynamic job prioritisation, configurable parameters, extensive fair share capabilities and backfill scheduling. Maui functions in an iterative manner like most job schedulers. It starts a new iteration when one of the following conditions is met: (1) a job or resource state alters; (2) a reservation boundary event occurs; (3) an external command to resume scheduling is issued; (4) a configuration timer expires. In each iteration, Maui follows the below steps \cite{Prabhakaran2016}:
		\begin{enumerate}
			\item Obtain resource records from TORQUE;
			\item Fetch workload information from TORQUE;
			\item Update statistics;
			\item Refresh reservations;
			\item Select jobs that are eligible for priority scheduling;
			\item Prioritise eligible jobs;
			\item Schedule jobs by priority and create reservations;
			\item Backfill jobs.
		\end{enumerate}
		Despite an abundance of algorithms, only a few scheduling strategies are practically in use by job schedulers. Backfilling scheduling \cite{Mualem2001} allows jobs to take the reserved job slots if this action does not delay the start of other jobs having reserved the resources, thus allowing large parallel jobs to execute and avoiding resource underutilisation. Differently, \textit{Gang scheduling} \cite{10.1007/11407522_1} attempts to take care of the situations when the runtime of a job is unknown, allowing smaller jobs to get fairer access to the resources. Both scheduling strategies are also seen in SLURM and backfilling can be also found in LSF.

		\subsubsection{SLURM}\label{subsec:slurm}
			\begin{figure}[!t]
			\centering
			\includegraphics[width=0.5\textwidth]{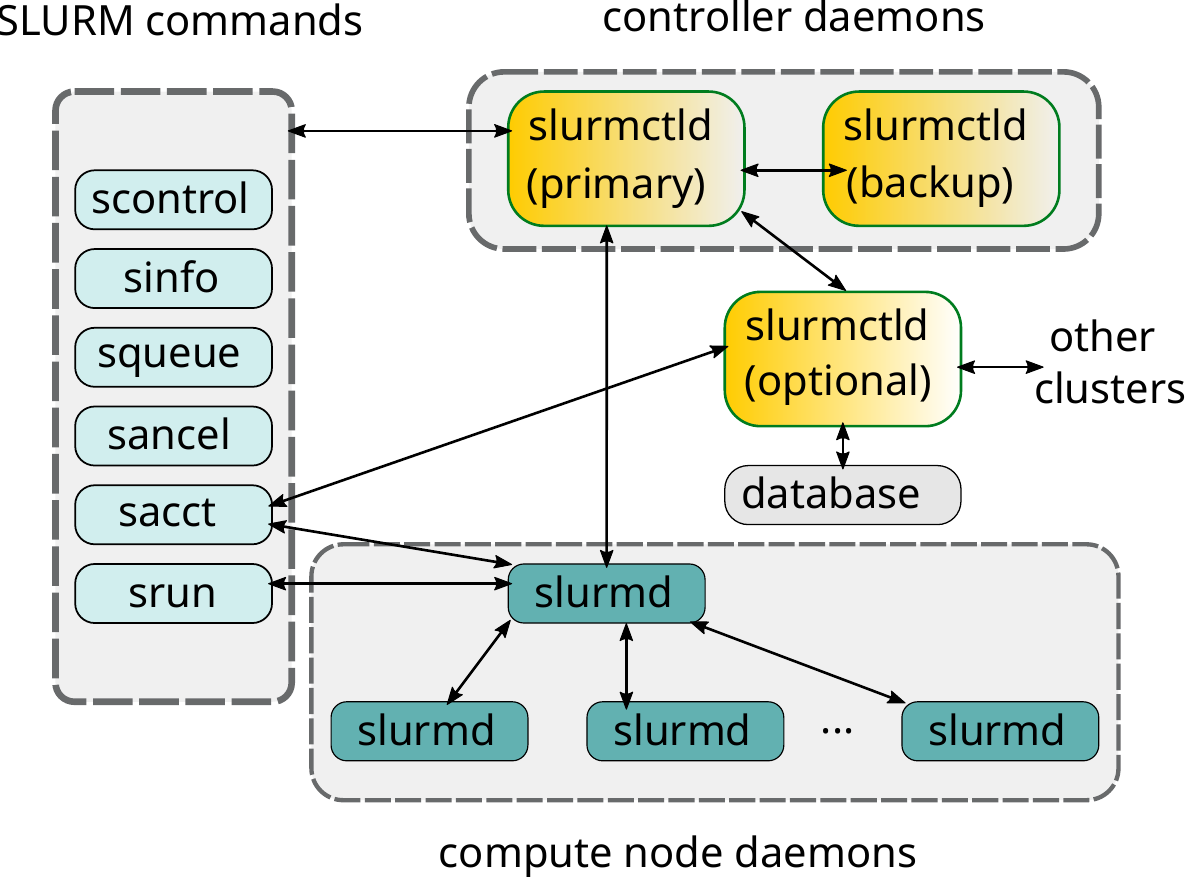}	
			\caption[]{SLURM structure.} 
			\label{fig:SlurmStructure}
		\end{figure}
		The structure of a SLURM (Simple Linux Utility for Resource Management) \cite{Jette02slurm:simple} managed cluster is composed of one or two SLURM servers and many compute nodes. Its procedure of job submission is similar to that of TORQUE. Fig.~\ref{fig:SlurmStructure} illustrates the structure of SLURM. Its server hosts the \texttt{slurmctld} daemon which is responsible for cluster resource and job management. SLURM servers and the corresponding \texttt{slurmctld} daemons can be deployed in an active/passive mode in order to provide services of high reliability for computing clusters. Each compute node hosts one instance of the \texttt{slurmd} daemon, which is responsible for job staging and execution. There are additional daemons, \textit{e.g.} \texttt{slurmdbd} which allows to collect and record accounting information for multiple SLURM-managed clusters and \texttt{slurmrestd} that can be used to interact with SLURM through a REST API (RESTful Application Programming Interface). The SLURM resource list is held as a part of the \texttt{slurm.conf} file located on SLURM server nodes, which contains a list of nodes including features (\textit{e.g.} CPU speed and model, amount of memory) and configured \textit{partitions} (named \textit{queue} in PBS) including partition names, list of associated nodes and job priority.

		Both PBS and SLURM have little (if at all) dedicated supports for container workloads. Containers are only scheduled as conventional HPC workloads, \textit{e.g} lacking of load-balancing supports.

		\subsubsection{Spectrum LSF}\label{subsec:spectrumLSF}
		
\begin{figure}[!t]
			\centering
			\includegraphics[width=0.5\textwidth]{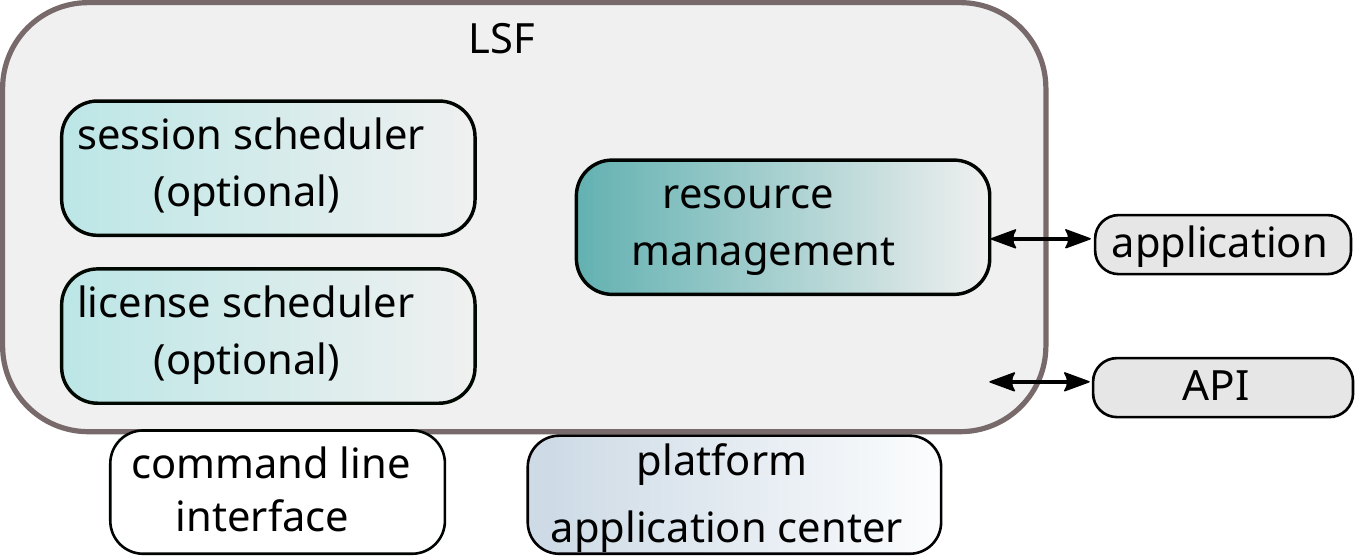}	
			\caption[]{Spectrum LSF structure.} 
			\label{fig:lsftructure}
		\end{figure}

		IBM platform Load Sharing Facility (LSF), targeted for enterprises, is designed for distributed HPC deployments. LSF is based on the Utopia job scheduler \cite{10.1002/spe.4380231203} developed at the University of Toronto. Its Session Scheduler runs and manages short-duration batch jobs, which enables users to submit multiple tasks as a single LSF job, consequently reduces the number of job scheduling decisions. Session Scheduler can efficiently share resources regardless of job execution time and can make thousands of scheduling decisions per second. These capabilities create a focus on throughput which is often critical for HPC workloads. Fig.~\ref{fig:lsftructure} illustrates the structure of LSF. Its license scheduler allows to make policies that control the way software licenses are shared among users within an organisation. Jobs are submitted via the command line interface, API or IBM platform application centre. Job submission carries similar procedure as in TORQUE.

		LSF supports container workloads: Docker, Singularity and Shifter. LSF configures container runtime control in the \textit{application profile}\footnote{LSF application profile: it is used to refine queue-level settings, or to exclude some jobs from queue-level parameters.} that is managed by the system administrator. Users do not need to consider which containers are used for their jobs, instead only need to submit their jobs to the application profile and LSF automatically manages the container runtime control. Section~\ref{subsubsec:cohabitation} elaborates this feature in more details.
		
		\subsection{Orchestration Frameworks on Cloud}\label{subsec:orch_cloud}
		Cloud clusters often include orchestration mechanisms to coordinate tasks and hardware resources. Cloud has evolved mature orchestrators to manage containers efficiently. Container orchestrators can offer \cite{Casalicchio2019, 10.5555/3175917, doi:10.1002/cpe.5668}:
		
		\begin{itemize}
			\item Resource limit control. Reserve a specific amount of CPUs and memory for a container, which restrains interference from other containers and provides information for scheduling decisions;
			\item Scheduling. It determines the policies that optimise the placement of containers on nodes;
			\item Load balancing. It distributes workloads among container instances;
			\item Health check. It verifies if a faulty container needs to be destroyed or replaced;
			\item Fault tolerance. It allows to maintain a desired number of containers;
			\item Auto-scaling. It automatically adds and removes containers.
		\end{itemize}  
		
		Additionally, a container orchestrator should also simplify networking, enable service discovery and support continuous deployment \cite{8125559}.

		\subsubsection{Kubernetes}\label{subsec:kubernetes}
		
		Kubernetes originally developed by Google is among the most popular open-source container orchestrators, which has a rapidly growing community and ecosystem with numerous platforms being developed upon it. The architecture of Kubernetes comprises a master node and a set of worker nodes. Kubernetes runs containers inside \textit{pods} that are scheduled to run either on master or worker nodes. A \textit{pod} can include one or multiple containers. Kubernetes provides its services via \textit{deployments} that are created by submission of \textit{yaml} files. Inside a \textit{yaml} file, users can specify services and computation to perform on the cluster. A user \textit{deployment} can be performed either on the master node or the worker nodes.
		
		Kubernetes is based on a highly modular architecture which abstracts the underlying infrastructure and allows internal customisation, such as the deployment of software-defined networks or storage solutions. It also supports various big-data frameworks, such as Hadoop MapReduce \cite{6821458}, Spark \cite{Zaharia:2016:ASU:3013530.2934664} and Kafka \cite{Narkhede:2017:KDG:3175825}. Kubernetes incorporates a powerful set of tools to control the life cycle of applications, \textit{e.g.} parameterised redeployment in case of failures and state management. Furthermore, it supports software-defined infrastructures\footnote{Software-defined infrastructure (SDI) is the definition of computing infrastructure entirely under the control of software with no operator or human intervention. It operates independent of any hardware-specific dependencies and is programmatically extensible.} \cite{6798709} and resource disaggregation \cite{10.5555/3026877.3026897} by leveraging container-based deployments and particular \textit{drivers} (\textit{e.g.} Container Runtime Interface driver, Container Storage Interface driver and Container Network Interface driver) based on standardised interfaces. These interfaces enable the definition of abstractions for fine-grain control of computation, states and communication in multi-tenant Cloud environments along with optimal usage of the underlying hardware resources.

		Kubernetes incorporates a scheduling system that permits users to specify different schedulers for each job. The scheduling system makes the decisions based on two steps before the actual scheduling operations: 
		\begin{enumerate}
			\item Node filtering. The scheduler locates the node(s) that fit(s) the workload, \textit{e.g.} a \textit{pod} is specified with node affinity, therefore, only certain nodes can meet the affinity requirements or some nodes may not include enough CPU resources to serve the request. Normally the scheduler does not traverse the entire node list, instead it selects the one/ones detected first.
			\item Node priority calculation. The scheduler calculates a score for each node, and the highest scoring node will run that \textit{pod}.
			
		\end{enumerate} 
		
		Kubernetes has started being utilised to assist HPC systems in container orchestration (Section~\ref{sec:stateart}).

		\subsubsection{Docker Swarm}
		Docker Swarm \cite{soppelsa2016native} is built for the Docker engine. It is a much simpler orchestrator comparing with Kubernetes, \textit{e.g.} it offers less rich functionalities, limited customisations and extensions. Docker Swarm is hence lightweight and suitable for small workloads. In contrast, Kubernetes is heavyweight for individual developers who may only want to set up an orchestrator for simplistic applications and perform infrequent deployments. Nevertheless, Docker Swarm still has its own API, and provides filtering, scheduling and load-balancing. API is a strong feature commonly used in Cloud orchestrators, as it enables applications or services to talk to each other and provides connections with other orchestrators.
		
		The functionalities of Docker Swarm may be applied to perform container orchestration on HPC systems as detailed in Section~\ref{subsubsec:cohabitation}. 
		
\subsubsection{Apache Mesos and YARN}\label{subsubsec:mesos}
		Apache Mesos \cite{10.5555/1972457.1972488} is a cluster manager that provides efficient resource isolation and sharing across distributed applications or frameworks. Mesos removes the centralised scheduling model that would otherwise require to compute global schedules for all the tasks running on the different frameworks connected to Mesos. Instead, each framework on a Mesos cluster can define its own scheduling strategies. For instance, Mesos can be connected with MPI or Hadoop \cite{10.5555/2285539}. Mesos utilises a master process to manage slave daemons running on each node. A typical Mesos cluster includes 3 $\sim$ 5 masters with one acting as the leader and the rest on standby. The master controls scheduling across frameworks through \textit{resource offers} that provide resource availability of the cluster to slaves. However, the master process only suggests the amount of resources that can be given to each framework according to the policies of organisations, \textit{e.g} fair sharing. Each framework rules which resources or tasks to accept. Once a \textit{resource offer} is accepted by a framework, the framework passes Mesos a description of the tasks. The slave comprises two components, \textit{i.e.} a scheduler registered to the master to receive resources and an executor process to run tasks from the frameworks.

		Mesos is a non-monolithic scheduler which acts as an arbiter that allocates resources across multiple schedulers, resolves conflicts, and ensures fair distribution of resources. Apache YARN (Yet Another Resource Negotiator) \cite{10.1145/2523616.2523633} is a monolithic scheduler which was developed in the first place to schedule Hadoop jobs. YARN is designed for long-running batch jobs and is unsuitable for long-running services and short-lived interactive queries.

		Mesosphere Marathon\footnote{\url{https://mesosphere.github.io/marathon/}} is a container orchestration framework for Apache Mesos.  Literature has seen the usage of Mesos together with Marathon in container orchestration on HPC systems as detailed in Section~\ref{subsubsec:cohabitation}. 
		
		\subsubsection{Ansible}
		Ansible \cite{10.5555/3125873} is a popular software orchestration tool. More specifically, it handles configuration management, application deployment, cloud provisioning, ad-hoc task execution, network automation and multi-node orchestration. The architecture of Ansible is simple and flexible, \textit{i.e.} it does not require a special server or daemons running on the nodes. Configurations are set by \textit{playbooks} that utilise \textit{yaml}
		to describe the \textit{automation jobs}, and connections to other nodes are via \texttt{ssh}. Nodes managed by Ansible are grouped into \textit{inventories} that can be defined by users or drawn from different Cloud environments. 
		
		Ansible is adopted by the SODALITE framework (Section~\ref{subsubsec:meta_orchestration}) as a key component to automatically build container images.
		
		\subsubsection{OpenStack} \label{subsubsec:openstack}
	
		OpenStack \cite{Sefraoui2012} is mostly deployed as infrastructure-as-a-service (IaaS)\footnote{IaaS offers resources such as compute, storage and network as services to users based on demand.} \cite{MANVI2014424} on Cloud. It can be utilised to deploy and manage cloud-based infrastructures that support various use cases, such as web hosting, big data projects, software as a service (SaaS) \cite{10.1007/978-3-540-74974-5_52} delivery and deployment of containers, VMs or bare-metal. It presents a scalable and highly adaptive open source architecture for Cloud solutions and helps to leverage hardware resources \cite{10.1145/2628194.2628195}. It also manages heterogeneous compute, storage and network resources.

		Together with its support of containers, container orchestrators such as Docker Swarm, Kubernetes and Mesos, Openstack enables the possibilities to quickly deploy, maintain, and upgrade complex and highly available infrastructures. OpenStack is also used in HPC communities to provide IaaS to end-users, enabling them to dynamically create isolated HPC environments.
		
		Academia and industry have developed a plethora of Cloud orchestrators. This article only reviews the ones that are mostly relevant to the HPC communities and the ones that have seen their usage in container orchestration for HPC systems, and the rest is out of the scope herein.

		\subsection{Bridge Orchestration Strategies Between HPC and Cloud}\label{sec:stateart}
		
\begin{figure*}[!t]
			\centering
			\includegraphics[width=0.85\textwidth]{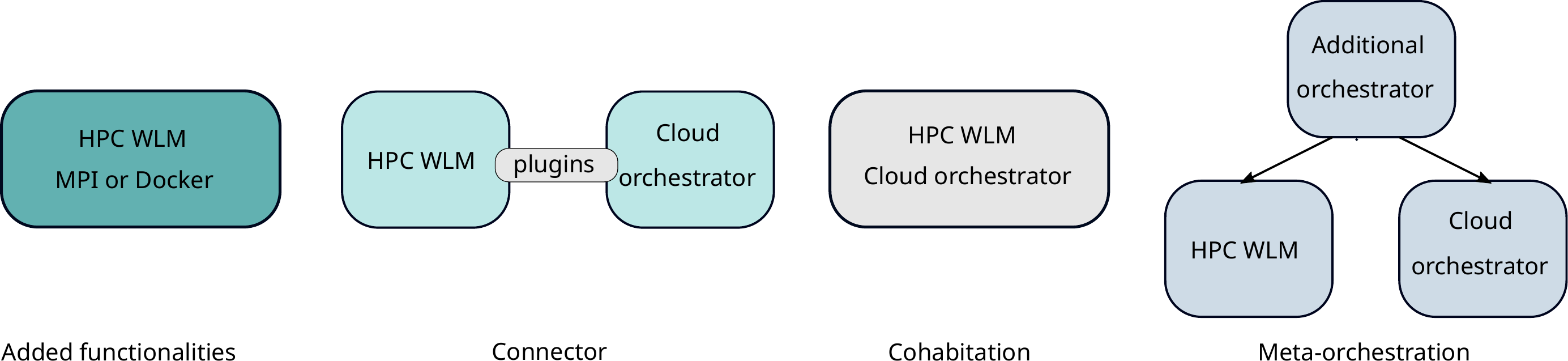}

			\caption{The four types of container orchestration on HPC systems. }\label{taxanomy_orchestration}
			
		\end{figure*}

		There are numerous works in literature \cite{10.1145/3297280.3297296, Casalicchio2019, Maenhaut2019, Buyya2019} on container orchestration for Cloud clusters, however, they are herein out of the scope. This section reviews the works that have been performed on the general issues of bridging the gap between conventional HPC and service-oriented infrastructures (Cloud). Overall, the state-of-the-art works on container orchestration for HPC systems fall into four categories as illustrated in Fig.~\ref{taxanomy_orchestration}.

		\begin{enumerate}	
			
			\item Added functionalities to HPC workload managers. It relies on workload managers for resource management and scheduling; meanwhile adopts additional software such as MPI for container orchestration.  
			
			\item Connector between Cloud and HPC. Containers are scheduled from Cloud clusters to HPC clusters. This architecture isolates the HPC resources from Cloud so as to ensure HPC environment security; meanwhile offers application developments with flexible environments and powerful computing resources. 
			
			\item Cohabitation. Workload managers and Cloud orchestrators co-exist on an HPC cluster, such as IBM LSF-Kubernetes. This gives a direction for the provision of HPC resources as services. In practice, the HPC workload managers and Cloud orchestrators do not coexist in one cluster.
			
			\item Meta-orchestration. An additional orchestrator is implemented on top of the Cloud orchestrator and HPC workload manager.
		\end{enumerate}

		There are pros and cons of the above four categories, which are outlined in Table~\ref{table:orchestration_hpc}. In addition, a research and engineering trend \cite{7868429, Saha2018, SOMASUNDARAM201447, 8919534, Evangelinos2008CloudCF} is to move HPC applications to Cloud, as Cloud provides flexible and cost-effective services which are favoured by small-sized or middle-sized business. \textit{Beltre etal.} \cite{8950981} proposed to manage HPC applications by Kubernetes on a Cloud cluster with powerful computing resources, \textit{e.g.} InfiniBand, which demonstrated comparable performance in containerised and bare-metal environments. The approach of this kind may be extended to HPC systems, however, remains unpractical for HPC centres to completely substitute their existing workload managers.

		\begin{table*}[ht]
			\caption[]{A list of the related work on container orchestration for HPC systems.}\label{table:orchestration_hpc}
			\centering  
		
				\tabulinesep=1.1mm
				\begin{tabu}to \textwidth{ X X X} 
					
					\toprule	
					\textbf{Orchestration approaches}	& \textbf{Advantages}& \textbf{Disadvantages}\\
					\toprule
					Added functionalities to WLM \cite{10.1145/3452370.3466001, 10.1145/2949550.2949562, 10.1007/978-3-319-20119-1_36} &Less intrusive  & Limited functionalities, security issues for usage of Docker on HPC\\ 	
				
					Connector between Cloud and HPC \cite{ Zhou2020, NaweiluoZhou2021, Zhou2021, zhoubook2021}  & Non-intrusive; flexible environments meanwhile powerful computing resources; exploit orchestration strategies of orchestration platforms & High network latency between Cloud and HPC   \\ 
					
					Cohabitation \cite{10.5555/3291656.3291707, Piras2019, 8360359, 10.1145/3019612.3019894, 10.1145/3019612.3019894, 8950981, 10.1145/2949550.2949562, 10.1007/978-3-319-20119-1_36}& Fully exploit the functionalities of orchestration platforms; flexible execution environments; enable HPC as services  & Intrusive, security issues \\ 
					
					Meta-orchestration \cite{8514380, 9356938, 9177340} & Less-intrusive; flexible environments meanwhile powerful computing resources; container orchestration strategies in addition to the ones given by Cloud orchestrator  & Increase architecture complexity; increase maintenance efforts \\ 
					
					\toprule		
				\end{tabu}

			\end{table*}

		\subsubsection{Added Functionalities to WLM}\label{subsubsec:added}   
		A potential research direction is to complement workload managers with container orchestration or make use of the existing HPC software stacks. Wofford \textit{et al.} \cite{10.1145/3452370.3466001} simply adopt Open Runtime Environment (orted) reference implementation from Open MPI to orchestrate container launch suitable for arbitrary batch schedulers.
		
		Julian \textit{et al.} \cite{10.1145/2949550.2949562} proposed their prototype for container orchestration in an HPC environment. A PBS-based HPC cluster can automatically scale up and down as load demands by launching Docker containers using the job scheduler Moab \cite{Moab2020}. Three containers serve as the front-end system, scheduler (it runs PBS and Moab inside) and compute node (launches $pbs\_mom$ daemon, see Section~\ref{subsec:torque}). More compute node containers are scheduled when there is no sufficient number of physical nodes. Unused containers are destroyed via external Python scripts when jobs complete. This approach may offer a solution for resource elasticity on HPC systems (Section~\ref{subsub:resouce_elasticity}). Similarly, an early study~\cite{10.1007/978-3-319-20119-1_36} described two models that can orchestrate Docker containers using an HPC workload manager. The former model launches a container to behave as one compute node which holds all assigned processes, whilst the latter boots a container per process by MPI launchers. The latter work seems to be outdated as to MPI applications which can be now automatically scaled with Singularity support.      
		
		\subsubsection{Connector between Cloud and HPC}
		
		Cloud technologies are evolving to be able to support complex applications of HPC, big data and AI. Nevertheless, the applications with intensive computation and high inter-processor communication could not scale well, particularly due to the lack of low latency networks (\textit{e.g.} InfiniBand) and the usage of network virtualisation for network isolation. A research and development trend is to converge HPC and Cloud in order to take advantage of the resource management and scheduling of both HPC and Cloud infrastructures with minimal intrusion to HPC environments. Furthermore, the software stack and workflows in Cloud and HPC are usually developed and maintained by different organisations and users with various goals and methodologies, hence a connector between HPC and Cloud systems would bridge the gap and solve compatibility problems. 
		
		Zhou \textit{et al.} \cite{Zhou2020, NaweiluoZhou2021, Zhou2021, zhoubook2021} described the design of a plugin named Torque-Operator that serves as the key component to its proposed hybrid architecture. The containerised AI applications are scheduled from the Kubernetes-managed Cloud cluster to the TORQUE-managed HPC cluster where the performance of the compute-intensive or data-demanding applications can be significantly enhanced. This approach is less intrusive to HPC systems, however, its architecture shows one drawback: the latency of the network bridging the Cloud and HPC clusters can be high, when a large amount of data needs to be transferred in-between.  
		
		DKube\footnote{\url{https://www.dkube.io/products/datascience/hpc-slurm.php}} is a commercial software that is able to execute a wide range of AI/ML components scheduled from Kubernetes to SLURM. The software comprises a Kubernetes plugin and a SLURM Plugin. The former is represented as a hub that runs MLOps (Machine Learning Operations) management and associated Kubernetes workloads, while the latter connects to SLURM.

		\subsubsection{Cohabitation}\label{subsubsec:cohabitation}

		Liu \textit{et al.} \cite{10.5555/3291656.3291707} showed how to dynamically migrate computing resources between HPC and OpenStack clusters based on demands. At a higher level, IBM has demonstrated the ability to run Kubernetes \textit{pods} on Spectrum LSF where LSF acts as a scheduler for Kubernetes. An additional Kubernetes scheduler daemon needs to be installed into the LSF cluster, which acts as a bridge between LSF and the Kuberentes server. \texttt{Kubelet} will execute and manage \textit{pod} lifecycle on target nodes in the normal fashion. IBM released LSF connector to Kubernetes, which makes use of the core LSF scheduling technologies and Kubernetes API functionalities. Kubernetes needs to be installed in a subset of the LSF managed HPC cluster. This architecture allows users to run Kubernetes and HPC batch jobs on the same infrastructure. The LSF scheduler is packed into containers and users submit jobs via \texttt{kubectl}. The LSF scheduler listens to the Kubernetes API server and translates \textit{pod} requests into jobs for the LSF scheduler. This approach can add additional heavy workloads to HPC systems, as Kubernetes relies deployments of services across clusters to perform load balancing, scheduling, auto scheduling, \textit{etc}.

		Piras \textit{et al.} \cite{Piras2019} implemented a method that expanded Kubernetes clusters with HPC clusters through Grid Engine. Submission is performed by PBS jobs to launch Kubernetes jobs. Therefore, HPC nodes are added to Kubernetes clusters by installing Kubernetes core components (\textit{i.e.} \texttt{kubeadm} and \texttt{Kubelet}) and Docker container engine. On HPC, especially HPC production systems in HPC centres, adding new software packages that require using root privileges can cause security risks and alter the working environments of current users. The security issues will be further elaborated in Section~\ref{subsubsec:security}.     
		
		Khan \textit{et al.} \cite{8360359} proposed to containerise HPC workloads and install Mesos and Marathon (Section~\ref{subsubsec:mesos}) on HPC clusters for resource management and container orchestration. Its orchestration system can obtain the appropriate resources satisfying the needs of requested services within defined Quality-of-Service (QoS) parameters, which is considered to be self-organised and self-managed meaning that users do not need to specifically request resource reservation. Nevertheless, this study has not shown insight into novel strategies of container orchestration for HPC systems.
		
		Wrede \textit{et al.} \cite{10.1145/3019612.3019894} performed their experiments on HPC clusters using Docker Swarm as the container orchestrator for automatic node scaling and using C++ algorithmic skeleton library Muesli \cite{Ciechanowicz2009} for load balance. Its proposed working environment is targeted for Cloud clusters. Usage of Docker cannot be easily extended to HPC infrastructures especially to HPC production systems due to the security risks. 
		\subsubsection{Meta-Orchestration}\label{subsubsec:meta_orchestration}
		Croupier \cite{8514380} is a plugin implemented on Cloudify\footnote{\url{https://cloudify.co/}} server that is located at a separate node in addition to the nodes that are managed by an HPC workload manager and a Cloud orchestrator. Croupier establishes a \textit{monitor} to collect the status of every infrastructure and the operations (\textit{e.g.} status of the HPC batch queue). Croupier together with Cloudify, can orchestrate batch applications in both HPC and Cloud environments. Similarly, Di Nitto \textit{et al.} \cite{9356938} presented the SODALITE\footnote{SODALITE: SOftware-Defined AppLication Infrastructures managemenT and Engineering. \url{https://www.sodalite.eu/}} framework by utilising XOpera\footnote{\url{https://github.com/xlab-si/xopera-opera}.} to manage the application deployment in heterogeneous infrastructures.
		
		Colonnelli \textit{et al.} \cite{9177340} presented a proof-of-concept framework (\textit{i.e.} Streamflow) to execute workflows on top of the hybrid architecture consisting of Kubernetes-managed Cloud and OCCAM \cite{Aldinucci2017} HPC cluster.

			\subsection{Section Highlights}\label{subsec:highlights}				
			
			HPC workload managers and Cloud orchestrators have distinct ways to manage clusters mainly because of their types of workloads and hardware resource availabilities. Table~\ref{table:comp_hpc_cloud_orchestrator} summaries the differences of key features between HPC workload managers and Cloud Orchestrators. Typical HPC jobs are large workloads with long but ascertainable execution time and large throughput. HPC jobs are often submitted to a batch queue within a workload manager where jobs wait to be scheduled from minutes to days. \textit{Per contra}, job requests can be granted immediately on Cloud as resources are available on demand. Batch-queuing is insufficient to satisfy the needs of Cloud communities: most of jobs are short in duration and the Cloud services are persistently long-running programs. Most of the HPC workload managers support \textit{Checkpointing} that allows applications to save the execution states of a running job and restart the job from the checkpointing when a crash happens. This feature is critical for an HPC application with execution time typically from hours to months. Because it enables the application to recover from error states or resume from the state when it was previously terminated by the workload manager when its walltime limit had been reached or resource allocation had been exceeded. In contrary, jobs on Cloud, which are often micro-service programs, are usually relaunched in case of failures \cite{REUTHER201876}. A container orchestrator offers an important property, \textit{i.e.} container status monitoring. This is practical for long-running Cloud services, as it can monitor and replace unhealthy containers per desired configuration. HPC systems do not offer the equivalence of container \textit{pod} which bundle performance monitoring services with the application itself as in Cloud systems \cite{Bernstein2014}. Additionally, HPC workload managers often do not provide capabilities of application elasticity or necessary API at execution time, however, these capabilities are important for task migration and resource allocation changes at runtime on Cloud \cite{8752819}.
			
			Section~\ref{sec:stateart} has reviewed the approaches to address the issues of container orchestration on HPC systems, which are summarised in Table~\ref{table:orchestration_hpc}. Overall, a container orchestrator on its own does not address all the requirements of HPC systems \cite{8457916}, as a result cannot replace existing workload managers in HPC centres. An HPC workload manager lacks micro-service support and deeply-integrated container management capabilities in which container orchestrators manifest their efficiency.

			\section{Research Challenges and Vision}\label{sec:vision_challenge} 
			
			The distinctions between Cloud and HPC clusters are diminishing, especially with the trend of HPC Clouds in industry \cite{10.1145/2425676.2425692}. HPC Cloud is becoming an alternative to on-premise HPC clusters for executing scientific applications and business analytics models \cite{10.1145/3150224}. Containerisation technologies help to ease the efforts of moving applications between Cloud and HPC. Nevertheless, not all applications are suitable for containerisation. For instance, in the typical HPC applications such as weather forecast or modelling of computational fluid dynamics, any virtualisation or high-latency networks can become the bottlenecks for performance. Containerisation in HPC still faces challenges of different folds (Section~\ref{subsec:challenges}).
			
			Interest in using containers on HPC systems is mainly due to the encapsulation and portability that yet may trade off with performance. In practice, containers deployed on HPC clusters often have large image size and as a result each HPC node can only host a few containers that are CPU-intensive and memory-demanding. In addition, implementation of AI frameworks such as TensorFlow and PyTorch \cite{DBLP:conf/nips/PaszkeGMLBCKLGA19} typically also have large container image size. Architecture of HPC containers should be able to easily integrate seamlessly with HPC workload managers. The research directions (Section~\ref{subsec:vision}) which can be envisioned are not only to adapt the existing functionalities from Cloud to HPC, but to also explore the potentials of containerisation so as to improve the current HPC systems and applications.

			\subsection{Challenges and Open Issues}\label{subsec:challenges}
			
			Although containerisation enables compatibility, portability and reproducibility, containerised environments still need to match the host architecture and exploit the underlying hardware. The challenges that containerisation faces on HPC systems are in three-fold: compatibility, security and performance. Some issues still remain as open questions. Table~\ref{table:challenge_solution} summarises the potential solutions to the research challenges and the open questions that will be discussed in this section.

			\begin{table*}[ht]
			\caption[]{Overview of research challenges and potential solutions.}\label{table:challenge_solution}
			\centering  
				\tabulinesep=1.0mm		
				\begin{tabu}to \textwidth {l |l |X| X }					
					\toprule						
			\multicolumn{2}{l|}{	\textbf{Research challenges}}	& \textbf{Potential solutions}  & \textbf{Open Questions}\\ 
						\toprule 

\multirow{3}{*}{Compatibility issues} &Library compatibility  &  OS updates,low-level container runtime libraries & \multirow{3}{*}{Reuse container images across platforms}\\ \cline{2-3}

		                 & Compatibility of engines and images & Container standardisation (e.g. OCI)\\ \cline{2-3}
			
			                       & Kernel optimisation& Using OS kernel to be library OS  \\ 

					\hline 
				\multicolumn{2}{l|}{	Security issues}& Private container registry, namespace settings, OS updates, rootless installation of container engines, avoid root processes inside containers  & Risk of using namespaces \\  
					\hline	
							
				\multicolumn{2}{l|}{	Performance degradation}& Trade-off between performance and portability &  Leverage hardware resources without losing portability  \\ 
					\toprule		
				\end{tabu}
				\\[1ex] 				
			\end{table*}

				\subsubsection{Library Compatibility Issues}
				Mapping container libraries and their dependencies to the host libraries can cause incompatibility. Glibc \cite{10.5555/1538674}, which is an implementation of C standard library that provides core supports and interfaces to kernel features, can be a common library dependency. The version of Glibc on the host may be older or newer than the one in the container image, consequently introducing symbol mismatches. Additionally, when the container OS (\textit{e.g.} Ubuntu 18.04) and the host OS are different (\textit{e.g.} CentOS 7), it is likely that some kernel ABI are incompatible, which may lead to container crashes or abnormal behaviours. This issue can also occur to MPI applications. As a result users must either build an exact version of the host MPI or have the privilege to mount the host MPI dependency path into the container.     
				
				A research direction to handle library mismatches between container images and hosts is to implement a container runtime library at a lower level. For instance, Nvidia implemented the library \texttt{libnvidia-container}\footnote{\url{https://github.com/NVIDIA/libnvidia-container}} that manages driver or library matching at container runtime, \textit{i.e.} using a hook interface to inject and/or activate the correct library versions. However, the \texttt{libnvidia-container} library can be only applied to Nividia GPUs. A significant modification of this library code is likely to be needed in order to be adapted for other GPU suppliers. In practice, such a compatibility layer would also require supports from different HPC interconnect and accelerator vendors.   
				
				\subsubsection{Compatibility Issues of Container Engines and Images}\label{subsubsec:issue_runtime_image}
				Not all Docker images can be converted by HPC container engines to their own formats. Moreover, to reuse HPC container implementations between container engines, users need to learn different container command lines to build the corresponding images, which further complicates adoption of containers for HPC applications. 
				
				This issue calls for container standardisation. OCI is a Linux foundation project that designs open standards for container image formats (a \texttt{filesystem bundle} or \texttt{rootfs}) and \textit{multiple data volume} \cite{7980161}. Some guidelines were proposed in \cite{8125559} , \textit{i.e.} a container should be:
				
				\begin{itemize}
					\item Not bound to higher-level frameworks, \textit{e.g.} an orchestration stack;
					\item Not tightly associated with any particular vendor or project;
					\item Portable across a wide variety of OSs, hardware, CPUs, clusters, \textit{etc}.
				\end{itemize}
				Unfortunately, this standard cannot guarantee that the runtime hooks built for one runtime can be used by another. For example, container privileges (\textit{e.g.} mount host filesystems) assumed by one container runtime may not be translated to unprivileged runtimes (\textit{e.g.} not all HPC centres have \texttt{mount} namespace enabled) \cite{8950982}.

				\subsubsection{Kernel Optimisation}
				In general, containers are forbidden by the host to install their own kernel modules for the purpose of application isolation \cite{10.1145/3297858.3304016}. This is a limitation for the applications requiring kernel customisation, because the kernels of their HPC hosts cannot be tuned and optimised. Shen \textit{et al.} \cite{10.1145/3297858.3304016} proposed an Xcontainer to address this issue by tuning the Linux kernel into library OS that supports binary compatibility. This functionality is yet to be explored in HPC containers. 
				
				\subsubsection{Security Issues}\label{subsubsec:security}
				Containers face three major threats \cite{10.1007/978-3-319-46079-6_48}:
				\begin{itemize}
					\item Privilege Escalation.  Attackers gain access to hosts and other containers by breaking out of their current containers.
					\item Denial-of-Service (DoS). An attack causes services to become inaccessible to users by disruption of a machine or network resources. 
					\item Information Leak. Confidential details of other containers are leaked and utilised for further attacks.
				\end{itemize}

				Multiple or many containers share a host kernel, therefore, one container may infect other containers. In this case, a container does not reduce attack surfaces, but rather brings multiple instances of attack surfaces. For example, starting from version V3.0, Singularity has added \texttt{Cgroups} support that allows users to limit the resources consumed by containers without the help from a batch scheduling system (\textit{e.g.} TORQUE). This feature helps to prevent DoS attacks when a container seizes control of all available system resources which prohibits other containers from operating properly.   
				
				Execution of HPC containers (including the Docker Engine starting from v19.03) does not require root privileges on the host. Containers in general adopt namespaces to isolate resources among users and map a root user inside a container to a non-root user on the host. The \texttt{User} namespace nevertheless is not a panacea to resolve all problems of resource isolation. \texttt{User} exposes code in the kernel to non-privileged users, which was previously limited to root users. A container environment is generated by users, and it is likely that some software inside a container may be embedded with security vulnerabilities. Root users inside a container may escalate their privileges via application level vulnerability. This can bring security issues to the kernel that does not account for mapped PIDs/GIDs. This issue can be addressed in two ways: (1) avoiding root processes inside HPC containers; (2) installing container engines with user permission instead of \texttt{sudo} installation. Security issues of the \texttt{user} namespace continue to be discovered even in the latest version of Linux kernels. Therefore, many HPC production centres have disabled the configuration of this namespace, which prevents usage of almost any state-of-the-art HPC containers. How to address the risks of using namespaces still remains an open question.

				\subsubsection{Performance Degradation}
				GPU and accelerators often require customised or proprietary libraries that need to be bound to container images so as to leverage performance. This operation is at the cost of portability \cite{8950982}. It is \textit{de facto} standard to utilise the optimised MPI libraries for HPC interconnects, such as InfiniBand and Slingshot \cite{10.1145/3491418.3530773}, and it is likely that the container performance degrades in a different HPC infrastructure \cite{8820966} (see Section~\ref{subsubsec:performance}). There is no simple solution to address this issue.  
				
				Another example presented in \cite{7573827} identified the performance loss due to increasing communication cost of MPI processes. This occurs when the number of containers (MPI processes running inside containers) rises on a single node, \textit{e.g.} point to point communication (\texttt{MPI\_Irecv, MPI\_Isend}), polling of pending asynchronous messages (\texttt{MPI\_Test}) and collective communication (\texttt{MPI\_Allreduce}).

				\subsection{Research and Engineering Opportunities}\label{subsec:vision}
				Research studies should continue working on solutions to the open question identified in Section~\ref{subsec:challenges}. This section discusses current research and engineering directions that are interesting, yet still need further development. This section also identifies new research opportunities that yet need to be explored. The presentation of this section is arranged from short-term vision to long-term efforts. Table~\ref{table:future_direction} summarises the potentials discovered in literature and the prospects given by the authors.

				\begin{table*}[ht]	
					\caption[]{Future directions of research and engineering.}\label{table:future_direction}
					\centering  

							\tabulinesep=0.8mm						
							\begin{tabu}to \textwidth{ l| X| X| X} 
							
								\toprule
								
								\textbf{Topics} &\textbf{Importance}&\textbf{State-of-art trends}& \textbf{Prospects given by the authors}\\
							\toprule
								Containerisation of AI in HPC&\makecell[tl]{Leverage HPC systems \\for ML/DL training }& \makecell[tl]{Containerised AI\\apps \& frameworks}	& \makecell[tl]{Improve scalability;\\ Enable out-of-box usage}   \\ 
								
								\hline
								HPC container registry & \makecell[tl]{Pre-build images accessible \\within HPC centres,\\ ensure container security}&\makecell[tl]{HPC centres set up\\private registries}& WLMs boot containers from registries without users awareness\\
								
							\hline
								Linux namespace guideline& Ensure security & HPC centres provide namespace guidelines& Different user groups to have different set of namespaces enabled\\
							
								\hline

								DevOps &Research reproducibility &Integration of Singularity with Jenkins &\makecell[tl]{HPC-specific DevOps tools}\\
								
							 \hline
						Middleware system &\makecell[tl]{Flexible and easy to plugin \\or plugout new components} &\makecell[tl]{Transfer Docker to HPC containers \\and perform the deployment\\ onto HPC systems} & Enable DevOps on HPC systems \\
								\hline
									Resource elasticity &\makecell[tl]{Flexible usage of hardware\\ resources} &\makecell[tl]{Kubernetes to instantiate the \\containerised HPC \\schedulers} & Integration of containers to introduce resource elasticity to WLM  \\
							\hline
								Moving toward minimal OS&\makecell[tl]{Reduce maintenance\\ efforts}& --&Maintain minimal OS kernel and containerised the rest of the HPC software stack \\			
								\bottomrule	
								\hline								
								\end{tabu}								
					\end{table*}

					\subsubsection{Containerisation of AI in HPC}\label{subsubsec:containerisation_AI}
					Model training of AI/DL applications can immensely benefit from the compute power (GPU or CPU), storage and security \cite{Mateescu2011} of HPC clusters in addition to the superior GPU-aware scheduling and features of workflow automation provided by workload managers. The trained models are subsequently deployed on Cloud for scalability at low cost and on HPC for computation speed. Exploiting HPC infrastructures for ML/DL training is becoming a topic of increasing importance \cite{Mayer2020}. For example, Fraunhofer\footnote{Fraunhofer: A German research organisation. \url{https://www.fraunhofer.de/}} has developed the software framework Carme\footnote{\url{https://www.itwm.fraunhofer.de/en/departments/hpc/data-analysis-and-machine-learning/carme-softwarestack.html}} that combines established open source ML and Data Science tools with HPC backends. The execution environments of the tools are provided by predefined Singularity containers.

					AI applications are usually developed with high-level scripting languages or frameworks, \textit{e.g.} TensorFlow and PyTorch, which often require connections to external systems to download a list of open-source software packages during execution. For instance, an AI application written in Python cannot be compiled into an executable that has included all the dependencies ready for execution as in C/C++. Therefore, the developers need flexibility to customise the execution environments. Since HPC environments, especially on HPC production systems, are often based on closed-source applications and their users have restricted account privileges and security restrictions \cite{8916576}, deployment of AI applications on HPC infrastructures is challenging. Besides the predefined module environments or virtual environments (such as Anaconda), containerisation can be an alternative candidate, which enables easy transition of AI workloads to HPC while fully taking advantage of HPC hardware and the optimised libraries of AI applications without compromising security. \textit{Huerta et al.} \cite{Huerta_2020} recommend three guidelines for containerisation of AI applications for HPC centres:
					\begin{itemize}
						\item Provide up-to-date documentation and tutorials to set up or launch containers.
						\item Maintain versatile and up-to-date base container images that users can clone and adapt, such as a container registry (see Section~\ref{subsubsec:Container_Registry}).
						\item Give instructions on installation or updates of software packages into containers. The AI program depends on distributed training software, such as Horovod \cite{Sergeev2018}, which then depends on system architecture and specific versions of software packages such as MPI.  
					\end{itemize}
					
					Increasing amount of new software frameworks are being developed using containerisation technologies to facilitate deployment of AI applications on HPC systems. Further research is still needed to improve scalability and enable out-of-box usage.   
					\subsubsection{HPC Container Registry}\label{subsubsec:Container_Registry}	
					Container registry is a useful repository to provide pre-built container images that can be accessed easily either by public or private users by pulling images to the host directly. It is portable to deploy applications in this way on Cloud clusters. Accesses to external networks are often blocked in HPC centres, so users need to upload images onto the clusters manually. One solution is to set up a private registry within the HPC centres that offer pre-built images suitable for the targeted systems and architectures. 
					
					A container registry is also a way to ensure container security. It is a good security practice to ensure that images executed on the HPC systems are signed and pulled from a trusted registry. Scanning vulnerabilities on the registry should be regularly performed. 
					
					To simplify usage, the future work can enable HPC workload managers to boot the default containers on the compute nodes (by pulling images from the private registry) which match the environments with all the required libraries and configuration files of user login nodes where users implement their own workflows and submit their jobs. The jobs should be started without user awareness of the presence of containers and without additional user intervention.

					\subsubsection{Linux Namespace Guidelines}\label{subsubsec:guideline}
					The set of Linux namespaces used within an implementation depends on the policies of HPC centres \cite{DBLP:journals/corr/abs-2007-10290}. HPC centres should provide clear instructions on the availabilities of namespaces. For example, different user groups may have different namespaces enabled or disabled. A minimal set of namespaces should be enabled for a general user group: \texttt{mount} and \texttt{user}, which are suitable for node-exclusive scheduling. \texttt{PID} and \texttt{Cgroups} should be provided to restrict resource usage and enforce process privacy, which are useful for shared-node scheduling. Advanced use cases may require additional sets of namespaces. When users submit the container jobs, workload managers can start the containers with appropriate namespaces enabled.

\subsubsection{DevOps}\label{subsubsec:devops}
DevOps aims at integrating efforts of development (Dev) and operations (Ops) to automate fast software delivery while ensuring correctness and reliability \cite{Huettermann2012, 10.1145/3359981}. This concept is influential in Cloud Computing and has been widely adopted in industry, as DevOps tools minimise the overhead of managing a large amount of micro-services. In HPC environments, typical applications have large workloads, hence the usage of DevOps should concentrate on research reproducibility. Nevertheless, the off-the-shelf DevOps tools are not well fitted for HPC environments, \textit{e.g.} the dependencies of MPI applications are too foreign for the state-of-the-art DevOps tools. A potential solution is to develop HPC-specific DevOps tools for the applications that are built and executed on on-premise clusters \cite{10.1145/3150224}. Unfortunately, HPC environments are known to be inflexible and typical HPC applications are optimised to leverage resources, thereby generation of DevOps workflows can be restricted and slow. Such obstacles can be overcome by containerisation, which may provision DevOps environments. For instance, \textit{Sampedro et al.} \cite{10.1145/3219104.3219147} integrate Singularity with Jenkins \cite{10.5555/3294743} that brings CICD\footnote{CICD: Continuous integration, delivery and deployment. It is widely used in DevOps communities.} practices into HPC workflows. Jenkins is an open-source automation platform for building and deploying software, which has been applied at some HPC sites as a general-purpose automation tool. 					

\subsubsection{Middleware System} \label{subsubsec:middleware}
A middleware system, which bridges container building environments with HPC resource managers and schedulers, can be flexible. A middleware system can be either located on an HPC cluster or connect to it with secured authentication. The main task of the middleware is to perform job deployment, job management, data staging and generating non-root container environments \cite{H_b_2020}. Different container engines can be swiftly switched, optimisation mechanisms can be adapted to the targeted HPC systems and workflow engines \cite{10.1145/3332301} can be easily plugged in. Middleware systems can be a future research direction that provides a portable way to enable DevOps in HPC centres.

\subsubsection{Resource Elasticity}\label{subsub:resouce_elasticity}
One major difference between resource management on HPC and Cloud is the elasticity \cite{9235080}, \textit{i.e.} an HPC workload manager runs on a fixed set of hardware resources and the workloads of its jobs at any point can not exceed the resource capacity, while Cloud orchestrators can scale up automatically the hardware resources to satisfy user needs (\textit{e.g.} AWS spot instances). Static reservation is a limitation for efficient resource usages on HPC systems \cite{10.1145/3555819.3555856}. One future direction of containerisation for HPC systems can work towards improvement of the elasticity of HPC infrastructure, which can be introduced to its workload manager. In \cite{9235080, greneche2022methodology}, the authors presented a novel architecture that utilises Kubernetes to instantiate the containerised HPC workload manager. In this way, the HPC infrastructure is dynamically instantiated on demand and can be served as a single-tenant or multi-tenant environment. A complete containerised environments on HPC system may be impractical and much more exploration is still needed. 

\subsubsection{Moving Towards Minimal OS}\label{subsubsec:minimal_os}
Containers may be utilised to partially substitute the current HPC software stack. Typical compute nodes on HPC clusters do not contain local storage (\textit{e.g.} hardware disk), therefore lose states after reboots. The compute node boots via a staged approach \cite{DBLP:journals/corr/abs-2007-10290}: (1) a kernel and initial RAM disk are loaded via a network device; (2) a root filesystem is mounted via the network. In a monolithic stateless system, modification of the software components often requires system rebooting to completely activate the functions of updates. Using containerised software packages on top of a minimal OS (base image) on the compute nodes, reduces the number of components in the kernel image, hence decreasing the frequency of node reboots. Furthermore, the base image of reduced size also simplifies the post-boot configurations that need to run in the OS image itself, consequently the node rebooting time is minimised. Additionally, when a failure occurs, a containerised service can be quickly replaced without affecting the entire system. Long-term research is required on HPC workload managers to control the software stack and workloads that are partially native and partially containerised. Moreover, it needs to explored whether containerisation of the entire OS on HPC systems is feasible.

\section{Concluding Remarks}\label{sec:conclude}
This paper presents a survey and taxonomy for the state-of-the-art container engines and container orchestration strategies specifically for HPC systems. It underlines differences of containerisation on Cloud and HPC systems. The research and engineering challenges are also discussed and the opportunities are envisioned. 

HPC systems start to utilise containers as thereof reduce environment complexity. Efforts have been also made to ameliorate container security on HPC systems. This article identified three points to increase the security level: (1) set on-site container registry, (2) give Linux namespaces guidelines (3) and remove root privilege meanwhile avoid permission escalation. Ideally, HPC containers should require no pre-installation of container engines or installation can be performed without root privileges, which not only meets the HPC security requirements but also simplifies the container usability.

Containers will continue to play a role in reducing the performance gap and deployment complexity between on-premise HPC clusters and public Clouds. Together with the advancement of low-latency networks and accelerators (\textit{e.g.} GPUs, TPUs \cite{10.1145/3079856.3080246}), it may eventually reshape the two fields. Containerised workloads can be moved from HPC to Cloud so as to temporarily relieve the peak demands and can be also scheduled from Cloud to HPC in order to exploit the powerful hardware resources. The research and engineering trend are working towards implementation of the present container orchestrators within HPC clusters, which however still remains experimental. Many studies have been devoted to container orchestration on Cloud, however, it can be foreseen that the strategies will be eventually introduced to HPC workload managers. 

In the future, it can be presumed that containerisation will play an essential role in application development, improve resource elasticity and reduce complexity of HPC software stacks.


%



\ifCLASSOPTIONcompsoc
  \section*{Acknowledgments}
\else
  \section*{Acknowledgment}
\fi

The project has received funding from the European Union’s Horizon 2020 research and innovation programme under grant agreement no 825355 (CYBELE), as well as through the project CATALYST funded by the Ministry of Science, Research and the Arts of the State of Baden-W\"urttemberg, Germany. \newline
\newline 
The authors would like to express the gratitude to Dr. Joseph Schuchart for proof-reading the contents.

\ifCLASSOPTIONcaptionsoff
  \newpage
\fi


\begin{thebibliography}{100}
	
	\bibitem{8360359}
	M.~{Khan}, T.~{Becker}, P.~{Kuppuudaiyar}, and A.~C. {Elster},
	``{Container-Based Virtualization for Heterogeneous HPC Clouds: Insights from
		the EU H2020 CloudLightning Project},'' in {\em 2018 IEEE International
		Conference on Cloud Engineering (IC2E)}, (Piscataway, New Jersey, US),
	pp.~392--397, IEEE, April 2018.
	
	\bibitem{DBLP:journals/spe/RodriguezB19}
	M.~A. Rodriguez and R.~Buyya, ``{Container-based cluster orchestration systems:
		A taxonomy and future directions},'' {\em Software: Practice and Experience},
	vol.~49, no.~5, pp.~698--719, 2019.
	
	\bibitem{8457916}
	L.~{Abdollahi Vayghan}, M.~A. {Saied}, M.~{Toeroe}, and F.~{Khendek},
	``{Deploying Microservice Based Applications with Kubernetes: Experiments and
		Lessons Learned},'' in {\em 2018 IEEE 11th International Conference on Cloud
		Computing (CLOUD)}, (Piscataway, New Jersey, US), pp.~970--973, IEEE, July
	2018.
	
	\bibitem{OlivierTerzo2022}
	J.~M. Olivier~Terzo, ed., {\em {HPC, Big Data, and AI Convergence Towards
			Exascale: Challenge and Vision}}.
	\newblock Boca Raton: CRC Press, 1st~ed., Jan. 2022.
	
	\bibitem{6114480}
	R.~McLay, K.~W. Schulz, W.~L. Barth, and T.~Minyard, ``{Best practices for the
		deployment and management of production HPC clusters},'' in {\em SC '11:
		Proceedings of 2011 International Conference for High Performance Computing,
		Networking, Storage and Analysis}, pp.~1--11, Nov 2011.
	
	\bibitem{8916576}
	D.~{Brayford}, S.~{Vallecorsa}, A.~{Atanasov}, F.~{Baruffa}, and W.~{Riviera},
	``{Deploying AI Frameworks on Secure HPC Systems with Containers},'' in {\em
		2019 IEEE High Performance Extreme Computing Conference (HPEC)}, (Piscataway,
	New Jersey, US), pp.~1--6, IEEE, Sep. 2019.
	
	\bibitem{Yi2019}
	G.~Yi and V.~Loia, ``{High-performance computing systems and applications for
		AI},'' {\em The Journal of Supercomputing}, vol.~75, 06 2019.
	
	\bibitem{10.4108/eai.25-10-2016.2266649}
	E.~Casalicchio, ``{Autonomic Orchestration of Containers: Problem Definition
		and Research Challenges},'' in {\em Proceedings of the 10th EAI International
		Conference on Performance Evaluation Methodologies and Tools on 10th EAI
		International Conference on Performance Evaluation Methodologies and Tools},
	VALUETOOLS¡¯16, (Brussels, BEL), p.~287šC290, ICST (Institute for Computer
	Sciences, Social-Informatics and Telecommunications Engineering), 2017.
	
	\bibitem{10.1109/CISIS.2015.35}
	A.~Tosatto, P.~Ruiu, and A.~Attanasio, ``{Container-Based Orchestration in
		Cloud: State of the Art and Challenges},'' in {\em Proceedings of the 2015
		Ninth International Conference on Complex, Intelligent, and Software
		Intensive Systems}, CISIS '15, (USA), p.~70šC75, IEEE Computer Society, 2015.
	
	\bibitem{Pahl2017}
	C.~Pahl, A.~Brogi, J.~Soldani, and P.~Jamshidi, ``{Cloud Container
		Technologies: a State-of-the-Art Review},'' {\em IEEE Transactions on Cloud
		Computing}, vol.~PP, pp.~1--1, 05 2017.
	
	\bibitem{Casalicchio2019}
	E.~Casalicchio, ``{Container Orchestration: A Survey},'' in {\em Systems
		Modeling: Methodologies and Tools} (A.~Puliafito and K.~S. Trivedi, eds.),
	pp.~221--235, Cham: Springer International Publishing, 2019.
	
	\bibitem{7868429}
	N.~{Nguyen} and D.~{Bein}, ``{Distributed MPI cluster with Docker Swarm
		mode},'' in {\em 2017 IEEE 7th Annual Computing and Communication Workshop
		and Conference (CCWC)}, pp.~1--7, Jan 2017.
	
	\bibitem{Bernstein2014}
	D.~{Bernstein}, ``{Containers and Cloud: From LXC to Docker to Kubernetes},''
	{\em IEEE Cloud Computing}, vol.~1, no.~3, pp.~81--84, 2014.
	
	\bibitem{10.1145/3378447}
	Z.~Zhong and R.~Buyya, ``{A Cost-Efficient Container Orchestration Strategy in
		Kubernetes-Based Cloud Computing Infrastructures with Heterogeneous
		Resources},'' {\em ACM Trans. Internet Technol.}, vol.~20, Apr. 2020.
	
	\bibitem{doi:10.1002/cpe.5668}
	E.~Casalicchio and S.~Iannucci, ``{The state-of-the-art in container
		technologies: Application, orchestration and security},'' {\em Concurrency
		and Computation: Practice and Experience}, vol.~n/a, no.~n/a, p.~e5668, 2019.
	\newblock e5668 cpe.5668.
	
	\bibitem{10.1145/3150224}
	M.~A.~S. Netto, R.~N. Calheiros, E.~R. Rodrigues, R.~L.~F. Cunha, and R.~Buyya,
	``{HPC Cloud for Scientific and Business Applications: Taxonomy, Vision, and
		Research Challenges},'' {\em ACM Comput. Surv.}, vol.~51, Jan. 2018.
	
	\bibitem{Merkel2014}
	D.~Merkel, ``{Docker: Lightweight Linux Containers for Consistent Development
		and Deployment},'' {\em Linux J.}, vol.~2014, pp.~76--90, Mar. 2014.
	
	\bibitem{7933304}
	J.~S. {Hale}, L.~{Li}, C.~N. {Richardson}, and G.~N. {Wells}, ``{Containers for
		Portable, Productive, and Performant Scientific Computing},'' {\em Computing
		in Science Engineering}, vol.~19, pp.~40--50, November 2017.
	
	\bibitem{10.1145/2988336.2988337}
	P.~Sharma, L.~Chaufournier, P.~Shenoy, and Y.~C. Tay, ``{Containers and Virtual
		Machines at Scale: A Comparative Study},'' in {\em Proceedings of the 17th
		International Middleware Conference}, Middleware '16, (New York, NY, USA),
	Association for Computing Machinery, 2016.
	
	\bibitem{7092949}
	R.~{Morabito}, J.~{Kj01llman}, and M.~{Komu}, ``{Hypervisors vs. Lightweight
		Virtualization: A Performance Comparison},'' in {\em 2015 IEEE International
		Conference on Cloud Engineering}, pp.~386--393, 2015.
	
	\bibitem{vmware2018}
	``{Containers on Virtual Machines or Bare Metals?},'' tech. rep., VMWare,
	VMware, Inc. 3401 Hillview Avenue Palo Alto CA 94304 USA, Dec. 2018.
	
	\bibitem{8820966}
	O.~Rudyy, M.~Garcia-Gasulla, F.~Mantovani, A.~Santiago, R.~Sirvent, and
	M.~Vš¢zquez, ``{Containers in HPC: A Scalability and Portability Study in
		Production Biological Simulations},'' in {\em 2019 IEEE International
		Parallel and Distributed Processing Symposium (IPDPS)}, pp.~567--577, May
	2019.
	
	\bibitem{Hu2019}
	G.~{Hu}, Y.~{Zhang}, and W.~{Chen}, ``{Exploring the Performance of Singularity
		for High Performance Computing Scenarios},'' in {\em 2019 IEEE 21st
		International Conference on High Performance Computing and Communications;
		IEEE 17th International Conference on Smart City; IEEE 5th International
		Conference on Data Science and Systems (HPCC/SmartCity/DSS)}, (Piscataway,
	New Jersey, US), pp.~2587--2593, IEEE, Aug 2019.
	
	\bibitem{Younge2017}
	A.~J. {Younge}, K.~{Pedretti}, R.~E. {Grant}, and R.~{Brightwell}, ``{A Tale of
		Two Systems: Using Containers to Deploy HPC Applications on Supercomputers
		and Clouds},'' in {\em 2017 IEEE International Conference on Cloud Computing
		Technology and Science (CloudCom)}, (Piscataway, New Jersey, US), pp.~74--81,
	IEEE, 2017.
	
	\bibitem{10.1145/3147213.3147231}
	J.~Zhang, X.~Lu, and D.~K. Panda, ``{Is Singularity-Based Container Technology
		Ready for Running MPI Applications on HPC Clouds?},'' in {\em Proceedings of
		The10th International Conference on Utility and Cloud Computing}, UCC 17,
	(New York, NY, USA), Association for Computing Machinery, 2017.
	
	\bibitem{10.1186/s13673-017-0124-3}
	J.~P. Martin, A.~Kandasamy, and K.~Chandrasekaran, ``{Exploring the Support for
		High Performance Applications in the Container Runtime Environment},'' {\em
		Hum.-Centric Comput. Inf. Sci.}, vol.~8, Dec. 2018.
	
	\bibitem{9284294}
	S.~Abraham, A.~K. Paul, R.~I.~S. Khan, and A.~R. Butt, ``{On the Use of
		Containers in High Performance Computing Environments},'' in {\em 2020 IEEE
		13th International Conference on Cloud Computing (CLOUD)}, pp.~284--293, Oct
	2020.
	
	\bibitem{10.1145/3126908.3126925}
	R.~Priedhorsky and T.~Randles, ``{Charliecloud: Unprivileged Containers for
		User-Defined Software Stacks in HPC},'' in {\em Proceedings of the
		International Conference for High Performance Computing, Networking, Storage
		and Analysis}, SC ¡¯17, (New York, NY, USA), Association for Computing
	Machinery, 2017.
	
	\bibitem{SenthilKumaran2017}
	S.~K. S., {\em {Practical LXC and LXD: Linux Containers for Virtualization and
			Orchestration}}.
	\newblock USA: Apress, 1st~ed., 2017.
	
	\bibitem{Saha2018}
	P.~Saha, A.~Beltre, P.~Uminski, and M.~Govindaraju, ``{Evaluation of Docker
		Containers for Scientific Workloads in the Cloud},'' in {\em Proceedings of
		the Practice and Experience on Advanced Research Computing}, PEARC 18, (New
	York, NY, USA), Association for Computing Machinery, 2018.
	
	\bibitem{Gerhardt_2017}
	L.~Gerhardt, W.~Bhimji, S.~Canon, M.~Fasel, D.~Jacobsen, M.~Mustafa, J.~Porter,
	and V.~Tsulaia, ``{Shifter: Containers for HPC},'' {\em Journal of Physics:
		Conference Series}, vol.~898, p.~082021, oct 2017.
	
	\bibitem{Gropp:1994:UMP:207387}
	W.~Gropp, E.~Lusk, and A.~Skjellum, {\em {Using MPI: Portable Parallel
			Programming with the Message-passing Interface}}.
	\newblock Cambridge, MA, USA: MIT Press, 1994.
	
	\bibitem{mpistandardv3.1}
	``{MPI: A Message-Passing Interface Standard}.''
	
	\bibitem{Kurtzer2017SingularitySC}
	G.~M. Kurtzer, V.~V. Sochat, and M.~Bauer, ``{Singularity: Scientific
		containers for mobility of compute},'' in {\em {PloS one}}, (San Francisco,
	California, United States), PLOS, 2017.
	
	\bibitem{10.1007/978-3-030-34356-9_5}
	L.~Benedicic, F.~A. Cruz, A.~Madonna, and K.~Mariotti, ``{Sarus: Highly
		Scalable Docker Containers for HPC Systems},'' in {\em High Performance
		Computing} (M.~Weiland, G.~Juckeland, S.~Alam, and H.~Jagode, eds.), (Cham),
	pp.~46--60, Springer International Publishing, 2019.
	
	\bibitem{10.1007/978-3-030-59851-8_23}
	H.~Gantikow, S.~Walter, and C.~Reich, ``{Rootless Containers with Podman for
		HPC},'' in {\em High Performance Computing} (H.~Jagode, H.~Anzt,
	G.~Juckeland, and H.~Ltaief, eds.), (Cham), pp.~343--354, Springer
	International Publishing, 2020.
	
	\bibitem{10.5555/3175917}
	K.~Hightower, B.~Burns, and J.~Beda, {\em {Kubernetes: Up and Running Dive into
			the Future of Infrastructure}}.
	\newblock Sebastopol, California, US: {O/'Reilly Media, Inc.}, 1st~ed., 2017.
	
	\bibitem{ruhela-TexasScholarWorks}
	A.~Ruhela, M.~Vaughn, S.~L. Harrell, G.~J. Zynda, J.~Fonner, R.~T. Evans, and
	T.~Minyard, ``{Containerization on Petascale HPC Clusters},'' Texas
	ScholarWorks, November 2020.
	
	\bibitem{7923813}
	A.~{Azab}, ``{Enabling Docker Containers for High-Performance and Many-Task
		Computing},'' in {\em 2017 IEEE International Conference on Cloud Engineering
		(IC2E)}, pp.~279--285, April 2017.
	
	\bibitem{Bahls2016}
	D.~Bahls, ``{Evaluating Shifter for HPC Applications},'' in {\em Cray User
		Group Conf.}, 2016.
	
	\bibitem{10.1007/978-3-030-34356-9_6}
	G.~Muscianisi, G.~Fiameni, and A.~Azab, ``{Singularity GPU Containers Execution
		on HPC Cluster},'' in {\em High Performance Computing} (M.~Weiland,
	G.~Juckeland, S.~Alam, and H.~Jagode, eds.), (Cham), pp.~61--68, Springer
	International Publishing, 2019.
	
	\bibitem{IMBbenchmark}
	IMB, ``{Introducing Intel MPI Benchmarks}.''
	\url{https://www.intel.com/content/www/us/en/developer/articles/technical/intel-mpi-benchmarks.html}
	(Sep, 2022).
	
	\bibitem{Dongarra2015}
	P.~L. Jack~Dongarra, Michael A.~Heroux, ``{HPCG Benchmark: a New Metric for
		Ranking High Performance Computing Systems},'' tech. rep., {Electrical
		Engineering and Computer Sciente Department}, 2015.
	
	\bibitem{crayxc40}
	{Cray XC 40}, ``{Overviw of Cray XC40 architecture}.''
	{https://www.alcf.anl.gov/files/CrayXC40Brochure.pdf} (Sep, 2022).
	
	\bibitem{doi:10.1137/1.9781611971811}
	J.~J. Dongarra, C.~B. Moler, J.~R. Bunch, and G.~W. Stewart, {\em {LINPACK
			Users' Guide}}.
	\newblock Society for Industrial and Applied Mathematics, 1979.
	
	\bibitem{NAMDbenchmark}
	{NAMD}, ``{Simulation for molecular dynamics}.''
	\url{https://www.ks.uiuc.edu/Research/namd/} (Sep,2022).
	
	\bibitem{VASPbenchmark}
	{VASP}, ``{Atomic scale materials modelling}.''
	\url{https://www.hpc.cineca.it/content/vasp-benchmark} (Sep. 2022).
	
	\bibitem{WRFbechmark}
	{WRF}, ``{Weather Research and Forecasting Model}.''
	\url{https://openbenchmarking.org/test/pts/wrf-1.0.0} (Sep,2022).
	
	\bibitem{AMBERbenchmark}
	{AMBER}, ``{Assisted Model Building with Energy Refinement}.''
	\url{http://ambermd.org/doc12/Amber18.pdf} (Sep, 2022).
	
	\bibitem{Abadi2016}
	M.~Abadi, P.~Barham, J.~Chen, Z.~Chen, A.~Davis, J.~Dean, M.~Devin,
	S.~Ghemawat, G.~Irving, M.~Isard, M.~Kudlur, J.~Levenberg, R.~Monga,
	S.~Moore, D.~G. Murray, B.~Steiner, P.~Tucker, V.~Vasudevan, P.~Warden,
	M.~Wicke, Y.~Yu, and X.~Zheng, ``{TensorFlow: A System for Large-scale
		Machine Learning},'' in {\em Proceedings of the 12th USENIX Conference on
		Operating Systems Design and Implementation}, OSDI'16, (Berkeley, CA, USA),
	pp.~265--283, USENIX Association, 2016.
	
	\bibitem{HPGMGbenchmark}
	{HPGMG}, ``{High-performance Geometric Multigrid ,Github}.''
	{https://github.com/hpgmg/hpgmg} (Sep, 2022).
	
	\bibitem{Hovestadt2003}
	M.~Hovestadt, O.~Kao, A.~Keller, and A.~Streit, ``{Scheduling in HPC Resource
		Management Systems: Queuing vs. Planning},'' in {\em Job Scheduling
		Strategies for Parallel Processing} (D.~Feitelson, L.~Rudolph, and
	U.~Schwiegelshohn, eds.), (Berlin, Heidelberg), pp.~1--20, Springer Berlin
	Heidelberg, 2003.
	
	\bibitem{Klusacek2015}
	D.~Klus\'{a}\v{c}ek, V.~Chlumsk\'{y}, and H.~Rudov\'{a}, ``Planning and
	optimization in torque resource manager,'' in {\em Proceedings of the 24th
		International Symposium on High-Performance Parallel and Distributed
		Computing}, (New York, NY, USA), Association for Computing Machinery, 2015.
	
	\bibitem{Staples2006}
	G.~Staples, ``Torque resource manager,'' in {\em Proceedings of the 2006
		ACM/IEEE Conference on Supercomputing}, (New York, NY, USA), p.~8,
	Association for Computing Machinery, 2006.
	
	\bibitem{10.1002/spe.4380231203}
	S.~Zhou, X.~Zheng, J.~Wang, and P.~Delisle, ``{Utopia: A Load Sharing Facility
		for Large, Heterogeneous Distributed Computer Systems},'' {\em Softw. Pract.
		Exper.}, vol.~23, p.~1305šC1336, Dec. 1993.
	
	\bibitem{923173}
	W.~Gentzsch, ``{Sun Grid Engine: towards creating a compute power grid},'' in
	{\em Proceedings First IEEE/ACM International Symposium on Cluster Computing
		and the Grid}, pp.~35--36, May 2001.
	
	\bibitem{10.5555/1169223.1169583}
	N.~Capit, G.~Da~Costa, Y.~Georgiou, G.~Huard, C.~Martin, G.~Mounie, P.~Neyron,
	and O.~Richard, ``{A Batch Scheduler with High Level Components},'' in {\em
		Proceedings of the Fifth IEEE International Symposium on Cluster Computing
		and the Grid (CCGrid'05) - Volume 2 - Volume 02}, CCGRID '05, (USA),
	p.~776šC783, IEEE Computer Society, 2005.
	
	\bibitem{Jette02slurm:simple}
	M.~A. Jette, A.~B. Yoo, and M.~Grondona, ``{SLURM: Simple Linux Utility for
		Resource Management},'' in {\em Proceedings of Job Scheduling Strategies for
		Parallel Processing (JSSPP)}, (publisher addr), pp.~44--60, Springer Berlin
	Heidelberg, 2003.
	
	\bibitem{10.1007/3-540-45540-X_6}
	D.~Jackson, Q.~Snell, and M.~Clement, ``{Core Algorithms of the Maui
		Scheduler},'' in {\em {Job Scheduling Strategies for Parallel Processing}}
	(D.~G. Feitelson and L.~Rudolph, eds.), (Berlin, Heidelberg), pp.~87--102,
	Springer Berlin Heidelberg, 2001.
	
	\bibitem{Prabhakaran2016}
	S.~Prabhakaran, {\em {Dynamic Resource Management and Job Scheduling for High
			Performance Computing}}.
	\newblock PhD thesis, Technische Universit{\"a}t Darmstadt, Darmstadt, August
	2016.
	
	\bibitem{Mualem2001}
	A.~W. {Mu'alem} and D.~G. {Feitelson}, ``{Utilization, predictability,
		workloads, and user runtime estimates in scheduling the IBM SP2 with
		backfilling},'' {\em IEEE Transactions on Parallel and Distributed Systems},
	vol.~12, no.~6, pp.~529--543, 2001.
	
	\bibitem{10.1007/11407522_1}
	D.~G. Feitelson, L.~Rudolph, and U.~Schwiegelshohn, ``{Parallel Job Scheduling
		--- A Status Report},'' in {\em Job Scheduling Strategies for Parallel
		Processing} (D.~G. Feitelson, L.~Rudolph, and U.~Schwiegelshohn, eds.),
	(Berlin, Heidelberg), pp.~1--16, Springer Berlin Heidelberg, 2005.
	
	\bibitem{8125559}
	A.~Khan, ``{Key Characteristics of a Container Orchestration Platform to Enable
		a Modern Application},'' {\em IEEE Cloud Computing}, vol.~4, pp.~42--48, Sep.
	2017.
	
	\bibitem{6821458}
	S.~{Pandey} and V.~{Tokekar}, ``{Prominence of MapReduce in Big Data
		Processing},'' in {\em 2014 Fourth International Conference on Communication
		Systems and Network Technologies}, (Piscataway, New Jersey, US),
	pp.~555--560, IEEE, 2014.
	
	\bibitem{Zaharia:2016:ASU:3013530.2934664}
	M.~Zaharia, R.~S. Xin, P.~Wendell, T.~Das, M.~Armbrust, A.~Dave, X.~Meng,
	J.~Rosen, S.~Venkataraman, M.~J. Franklin, A.~Ghodsi, J.~Gonzalez,
	S.~Shenker, and I.~Stoica, ``{Apache Spark: A Unified Engine for Big Data
		Processing},'' {\em {Commun. ACM}}, vol.~59, pp.~56--65, Oct. 2016.
	
	\bibitem{Narkhede:2017:KDG:3175825}
	N.~Narkhede, G.~Shapira, and T.~Palino, {\em {Kafka: The Definitive Guide
			Real-Time Data and Stream Processing at Scale}}.
	\newblock Sebastopol, California, US: O'Reilly Media, Inc., 1st~ed., 2017.
	
	\bibitem{6798709}
	G.~Kandiraju, H.~Franke, M.~D. Williams, M.~Steinder, and S.~M. Black,
	``{Software defined infrastructures},'' {\em IBM Journal of Research and
		Development}, vol.~58, pp.~2:1--2:13, March 2014.
	
	\bibitem{10.5555/3026877.3026897}
	P.~X. Gao, A.~Narayan, S.~Karandikar, J.~Carreira, S.~Han, R.~Agarwal,
	S.~Ratnasamy, and S.~Shenker, ``{Network Requirements for Resource
		Disaggregation},'' in {\em Proceedings of the 12th USENIX Conference on
		Operating Systems Design and Implementation}, OSDI¡¯16, (USA), p.~249šC264,
	USENIX Association, 2016.
	
	\bibitem{soppelsa2016native}
	F.~Soppelsa and C.~Kaewkasi, {\em Native docker clustering with swarm}.
	\newblock Packt Publishing Ltd, 2016.
	
	\bibitem{10.5555/1972457.1972488}
	B.~Hindman, A.~Konwinski, M.~Zaharia, A.~Ghodsi, A.~D. Joseph, R.~Katz,
	S.~Shenker, and I.~Stoica, ``{Mesos: A Platform for Fine-Grained Resource
		Sharing in the Data Center},'' in {\em {Proceedings of the 8th USENIX
			Conference on Networked Systems Design and Implementation}}, NSDI¡¯11, (USA),
	p.~295šC308, USENIX Association, 2011.
	
	\bibitem{10.5555/2285539}
	T.~White, {\em {Hadoop: The Definitive Guide}}.
	\newblock O¡¯Reilly Media, Inc., 2012.
	
	\bibitem{10.1145/2523616.2523633}
	V.~K. Vavilapalli, A.~C. Murthy, C.~Douglas, S.~Agarwal, M.~Konar, R.~Evans,
	T.~Graves, J.~Lowe, H.~Shah, S.~Seth, B.~Saha, C.~Curino, O.~O'Malley,
	S.~Radia, B.~Reed, and E.~Baldeschwieler, ``{Apache Hadoop YARN: Yet Another
		Resource Negotiator},'' in {\em Proceedings of the 4th Annual Symposium on
		Cloud Computing}, SOCC '13, (New York, NY, USA), Association for Computing
	Machinery, 2013.
	
	\bibitem{10.5555/3125873}
	G.~Sammons, {\em {Exploring Ansible 2: Fast and Easy Guide}}.
	\newblock North Charleston, SC, USA: CreateSpace Independent Publishing
	Platform, 2016.
	
	\bibitem{Sefraoui2012}
	O.~Sefraoui, M.~Aissaoui, and M.~Eleuldj, ``{OpenStack: Toward an Open-Source
		Solution for Cloud Computing},'' {\em International Journal of Computer
		Applications}, vol.~55, pp.~38--42, 10 2012.
	
	\bibitem{MANVI2014424}
	S.~S. Manvi and G.~K. Shyam, ``{Resource management for Infrastructure as a
		Service (IaaS) in cloud computing: A survey},'' {\em Journal of Network and
		Computer Applications}, vol.~41, pp.~424--440, 2014.
	
	\bibitem{10.1007/978-3-540-74974-5_52}
	W.~Sun, K.~Zhang, S.-K. Chen, X.~Zhang, and H.~Liang, ``{Software as a Service:
		An Integration Perspective},'' in {\em Service-Oriented Computing -- ICSOC
		2007} (B.~J. Kr{\"a}mer, K.-J. Lin, and P.~Narasimhan, eds.), (Berlin,
	Heidelberg), pp.~558--569, Springer Berlin Heidelberg, 2007.
	
	\bibitem{10.1145/2628194.2628195}
	T.~Rosado and J.~Bernardino, ``{An Overview of Openstack Architecture},'' in
	{\em Proceedings of the 18th International Database Engineering \&amp;
		Applications Symposium}, IDEAS '14, (New York, NY, USA), p.~366šC367,
	Association for Computing Machinery, 2014.
	
	\bibitem{10.1145/3297280.3297296}
	G.~P. Fernandez and A.~Brito, ``{Secure Container Orchestration in the Cloud:
		Policies and Implementation},'' in {\em Proceedings of the 34th ACM/SIGAPP
		Symposium on Applied Computing}, SAC ¡¯19, (New York, NY, USA), p.~138šC145,
	Association for Computing Machinery, 2019.
	
	\bibitem{Maenhaut2019}
	P.-J. Maenhaut, B.~Volckaert, V.~Ongenae, and F.~De~Turck, ``{Resource
		Management in a Containerized Cloud: Status and Challenges},'' {\em Journal
		of Network and Systems Management}, vol.~28, pp.~197--246, 11 2019.
	
	\bibitem{Buyya2019}
	R.~{Buyya} and S.~N. {Srirama}, {\em {A Lightweight Container Middleware for
			Edge Cloud Architectures}}, pp.~145--170.
	\newblock Wiley, 2019.
	
	\bibitem{SOMASUNDARAM201447}
	T.~S. Somasundaram and K.~Govindarajan, ``{CLOUDRB: A framework for scheduling
		and managing High-Performance Computing (HPC) applications in science
		cloud},'' {\em Future Generation Computer Systems}, vol.~34, pp.~47--65,
	2014.
	\newblock Special Section: Distributed Solutions for Ubiquitous Computing and
	Ambient Intelligence.
	
	\bibitem{8919534}
	K.~Cho, H.~Lee, K.~Bang, and S.~Kim, ``{Possibility of HPC Application on Cloud
		Infrastructure by Container Cluster},'' in {\em 2019 IEEE International
		Conference on Computational Science and Engineering (CSE) and IEEE
		International Conference on Embedded and Ubiquitous Computing (EUC)},
	pp.~266--271, Aug 2019.
	
	\bibitem{Evangelinos2008CloudCF}
	C.~Evangelinos and C.~Hill, ``{Cloud Computing for parallel Scientific HPC
		Applications: Feasibility of Running Coupled Atmosphere-Ocean Climate
		Modelson Amazon¡¯s EC2},'' 2008.
	
	\bibitem{8950981}
	A.~M. Beltre, P.~Saha, M.~Govindaraju, A.~Younge, and R.~E. Grant, ``{Enabling
		HPC Workloads on Cloud Infrastructure Using Kubernetes Container
		Orchestration Mechanisms},'' in {\em 2019 IEEE/ACM International Workshop on
		Containers and New Orchestration Paradigms for Isolated Environments in HPC
		(CANOPIE-HPC)}, (Los Alamitos, CA, USA), pp.~11--20, IEEE Computer Society,
	nov 2019.
	
	\bibitem{10.1145/3452370.3466001}
	Q.~Wofford, P.~G. Bridges, and P.~Widener, ``{A Layered Approach for Modular
		Container Construction and Orchestration in HPC Environments},'' in {\em
		Proceedings of the 11th Workshop on Scientific Cloud Computing}, ScienceCloud
	'21, (New York, NY, USA), p.~1šC8, Association for Computing Machinery, 2021.
	
	\bibitem{10.1145/2949550.2949562}
	S.~Julian, M.~Shuey, and S.~Cook, ``{Containers in Research: Initial
		Experiences with Lightweight Infrastructure},'' in {\em Proceedings of the
		XSEDE16 Conference on Diversity, Big Data, and Science at Scale}, XSEDE16,
	(New York, NY, USA), Association for Computing Machinery, 2016.
	
	\bibitem{10.1007/978-3-319-20119-1_36}
	J.~Higgins, V.~Holmes, and C.~Venters, ``{Orchestrating Docker Containers in
		the HPC Environment},'' in {\em {High Performance Computing}} (J.~M. Kunkel
	and T.~Ludwig, eds.), (Cham), pp.~506--513, Springer International
	Publishing, 2015.
	
	\bibitem{Zhou2020}
	N.~Zhou, Y.~Georgiou, L.~Zhong, H.~Zhou, and M.~Pospieszny, ``{Container
		Orchestration on HPC Systems},'' in {\em 2020 IEEE International Conference
		on Cloud Computing (CLOUD)}, (Piscataway, New Jersey, US), IEEE, 2020.
	
	\bibitem{NaweiluoZhou2021}
	N.~Zhou, Y.~Georgiou, M.~Pospieszny, L.~Zhong, H.~Zhou, C.~Niethammer,
	B.~Pejak, O.~Marko, and D.~Hoppe, ``{Container Orchestration on HPC Systems
		through Kubernetes},'' {\em {Journal of Cloud Computing: Advances, Systems
			and Applications}}, 2021.
	
	\bibitem{Zhou2021}
	N.~Zhou, ``{Containerization and Orchestration on HPC Systems},'' in {\em
		Sustained Simulation Performance 2019 and 2020}, Springer International
	Publishing, 2021.
	
	\bibitem{zhoubook2021}
	N.~Zhou, L.~Zhong, D.~Hoppe, B.~Pejak, O.~Marko, J.~Cardona, M.~Czerkawski,
	I.~Andonovic, C.~Michie, C.~Tachtatzis, E.~Alexakis, P.~Mavrepis,
	D.~Kyriazis, and M.~Pospieszny, {\em {CYBELE: A Hybrid Architecture for HPC
			and Big Data for AI applications in Agriculture}}, ch.~13.
	\newblock CRC press, 2022.
	
	\bibitem{10.5555/3291656.3291707}
	F.~Liu, K.~Keahey, P.~Riteau, and J.~Weissman, ``{Dynamically Negotiating
		Capacity between On-Demand and Batch Clusters},'' in {\em Proceedings of the
		International Conference for High Performance Computing, Networking, Storage,
		and Analysis}, SC ¡¯18, (Piscataway, New Jersey, United States), IEEE Press,
	2018.
	
	\bibitem{Piras2019}
	M.~E. Piras, L.~Pireddu, M.~Moro, and G.~Zanetti, ``{Container Orchestration on
		HPC Clusters},'' in {\em High Performance Computing} (M.~Weiland,
	G.~Juckeland, S.~Alam, and H.~Jagode, eds.), (Cham), pp.~25--35, Springer
	International Publishing, 2019.
	
	\bibitem{10.1145/3019612.3019894}
	F.~Wrede and V.~von Hof, ``{Enabling Efficient Use of Algorithmic Skeletons in
		Cloud Environments: Container-Based Virtualization for Hybrid CPU-GPU
		Execution of Data-Parallel Skeletons},'' in {\em Proceedings of the Symposium
		on Applied Computing}, SAC ¡¯17, (New York, NY, USA), p.~1593šC1596,
	Association for Computing Machinery, 2017.
	
	\bibitem{8514380}
	J.~Carnero and F.~J. Nieto, ``{Running Simulations in HPC and Cloud Resources
		by Implementing Enhanced TOSCA Workflows},'' in {\em 2018 International
		Conference on High Performance Computing Simulation (HPCS)}, pp.~431--438,
	July 2018.
	
	\bibitem{9356938}
	E.~Di~Nitto, J.~Gorro09ogoitia, I.~Kumara, G.~Meditskos, D.~Radolovi04,
	K.~Sivalingam, and R.~S. Gonzš¢lez, ``{An Approach to Support Automated
		Deployment of Applications on Heterogeneous Cloud-HPC Infrastructures},'' in
	{\em 2020 22nd International Symposium on Symbolic and Numeric Algorithms for
		Scientific Computing (SYNASC)}, pp.~133--140, Sep. 2020.
	
	\bibitem{9177340}
	I.~Colonnelli, B.~Cantalupo, I.~Merelli, and M.~Aldinucci, ``{StreamFlow:
		cross-breeding cloud with HPC},'' {\em IEEE Transactions on Emerging Topics
		in Computing}, pp.~1--1, aug 5555.
	
	\bibitem{Moab2020}
	``{Moab HPC Suite}.''
	https://support.adaptivecomputing.com/wp-content/uploads/2019/06/Moab-HPC-Suite\_datasheet\_20190611.pdf
	(Access on 08/07/2020).
	
	\bibitem{Ciechanowicz2009}
	P.~Ciechanowicz, M.~Poldner, and H.~Kuchen, ``{The M\"unster Skeleton Library
		Muesli: A comprehensive overview},'' tech. rep., University of Mš¹nster,
	European Research Center for Information Systems (ERCIS), 01 2009.
	
	\bibitem{Aldinucci2017}
	M.~Aldinucci, S.~Bagnasco, S.~Lusso, P.~Pasteris, S.~Rabellino, and S.~Vallero,
	``{OCCAM}: a flexible, multi-purpose and extendable {HPC} cluster,'' {\em
		Journal of Physics: conference series}, vol.~898, p.~082039, oct 2017.
	
	\bibitem{REUTHER201876}
	A.~Reuther, C.~Byun, W.~Arcand, D.~Bestor, B.~Bergeron, M.~Hubbell, M.~Jones,
	P.~Michaleas, A.~Prout, A.~Rosa, and J.~Kepner, ``{Scalable system scheduling
		for HPC and big data},'' {\em Journal of Parallel and Distributed Computing},
	vol.~111, pp.~76 -- 92, 2018.
	
	\bibitem{8752819}
	A.~{Souza}, M.~{Rezaei}, E.~{Laure}, and J.~{Tordsson}, ``{Hybrid Resource
		Management for HPC and Data Intensive Workloads},'' in {\em 2019 19th
		IEEE/ACM International Symposium on Cluster, Cloud and Grid Computing
		(CCGRID)}, pp.~399--409, May 2019.
	
	\bibitem{10.1145/2425676.2425692}
	P.~Marshall, H.~Tufo, and K.~Keahey, ``{High-Performance Computing and the
		Cloud: A Match Made in Heaven or Hell?},'' {\em XRDS}, vol.~19, p.~52šC57,
	Mar. 2013.
	
	\bibitem{DBLP:conf/nips/PaszkeGMLBCKLGA19}
	A.~Paszke, S.~Gross, F.~Massa, A.~Lerer, J.~Bradbury, G.~Chanan, T.~Killeen,
	Z.~Lin, N.~Gimelshein, L.~Antiga, A.~Desmaison, A.~K{\"{o}}pf, E.~Yang,
	Z.~DeVito, M.~Raison, A.~Tejani, S.~Chilamkurthy, B.~Steiner, L.~Fang,
	J.~Bai, and S.~Chintala, ``{PyTorch: An Imperative Style, High-Performance
		Deep Learning Library},'' in {\em Advances in Neural Information Processing
		Systems 32: Annual Conference on Neural Information Processing Systems 2019,
		NeurIPS 2019, 8-14 December 2019, Vancouver, BC, Canada} (H.~M. Wallach,
	H.~Larochelle, A.~Beygelzimer, F.~d'Alch{\'{e}}{-}Buc, E.~B. Fox, and
	R.~Garnett, eds.), pp.~8024--8035, 2019.
	
	\bibitem{10.5555/1538674}
	B.~Gough, {\em {GNU Scientific Library Reference Manual - Third Edition}}.
	\newblock Network Theory Ltd., 3rd~ed., 2009.
	
	\bibitem{7980161}
	S.~Nadgowda, S.~Suneja, N.~Bila, and C.~Isci, ``{Voyager: Complete Container
		State Migration},'' in {\em 2017 IEEE 37th International Conference on
		Distributed Computing Systems (ICDCS)}, pp.~2137--2142, June 2017.
	
	\bibitem{8950982}
	R.~S. Canon and A.~Younge, ``{A Case for Portability and Reproducibility of HPC
		Containers},'' in {\em 2019 IEEE/ACM International Workshop on Containers and
		New Orchestration Paradigms for Isolated Environments in HPC (CANOPIE-HPC)},
	pp.~49--54, Nov 2019.
	
	\bibitem{10.1145/3297858.3304016}
	Z.~Shen, Z.~Sun, G.-E. Sela, E.~Bagdasaryan, C.~Delimitrou, R.~Van~Renesse, and
	H.~Weatherspoon, ``{X-Containers: Breaking Down Barriers to Improve
		Performance and Isolation of Cloud-Native Containers},'' in {\em Proceedings
		of the Twenty-Fourth International Conference on Architectural Support for
		Programming Languages and Operating Systems}, ASPLOS '19, (New York, NY,
	USA), p.~121šC135, Association for Computing Machinery, 2019.
	
	\bibitem{10.1007/978-3-319-46079-6_48}
	H.~Gantikow, C.~Reich, M.~Knahl, and N.~Clarke, ``{Providing Security in
		Container-Based HPC Runtime Environments},'' in {\em High Performance
		Computing} (M.~Taufer, B.~Mohr, and J.~M. Kunkel, eds.), (Cham),
	pp.~685--695, Springer International Publishing, 2016.
	
	\bibitem{10.1145/3491418.3530773}
	K.~Shafie~Khorassani, C.~C. Chen, B.~Ramesh, A.~Shafi, H.~Subramoni, and
	D.~Panda, ``{High Performance MPI over the Slingshot Interconnect: Early
		Experiences},'' in {\em Practice and Experience in Advanced Research
		Computing}, PEARC '22, (New York, NY, USA), Association for Computing
	Machinery, 2022.
	
	\bibitem{7573827}
	J.~Zhang, X.~Lu, and D.~K. Panda, ``{High Performance MPI Library for
		Container-Based HPC Cloud on InfiniBand Clusters},'' in {\em 2016 45th
		International Conference on Parallel Processing (ICPP)}, pp.~268--277, Aug
	2016.
	
	\bibitem{Mateescu2011}
	G.~Mateescu, W.~Gentzsch, and C.~J. Ribbens, ``{Hybrid Computing-Where HPC
		Meets Grid and Cloud Computing},'' {\em Future Gener. Comput. Syst.},
	vol.~27, p.~440šC453, May 2011.
	
	\bibitem{Mayer2020}
	R.~Mayer and H.-A. Jacobsen, ``{Scalable Deep Learning on Distributed
		Infrastructures: Challenges, Techniques, and Tools},'' {\em {ACM Comput.
			Surv.}}, vol.~53, Feb. 2020.
	
	\bibitem{Huerta_2020}
	E.~A. Huerta, A.~Khan, E.~Davis, C.~Bushell, W.~D. Gropp, D.~S. Katz,
	V.~Kindratenko, S.~Koric, W.~T.~C. Kramer, B.~McGinty, and et~al.,
	``{Convergence of artificial intelligence and high performance computing on
		NSF-supported cyberinfrastructure},'' {\em Journal of Big Data}, vol.~7, Oct
	2020.
	
	\bibitem{Sergeev2018}
	A.~Sergeev and M.~D. Balso, ``{Horovod: fast and easy distributed deep learning
		in TensorFlow},'' {\em CoRR}, vol.~abs/1802.05799, 2018.
	
	\bibitem{DBLP:journals/corr/abs-2007-10290}
	B.~S. Allen, M.~A. Ezell, P.~Peltz, D.~Jacobsen, E.~Roman, C.~Lueninghoener,
	and J.~L. Wofford, ``{Modernizing the {HPC} System Software Stack},'' {\em
		CoRR}, vol.~abs/2007.10290, 2020.
	
	\bibitem{Huettermann2012}
	M.~H\"uttermann, {\em {DevOps for Developers}}.
	\newblock Apress, 2012.
	
	\bibitem{10.1145/3359981}
	L.~Leite, C.~Rocha, F.~Kon, D.~Milojicic, and P.~Meirelles, ``{A Survey of
		DevOps Concepts and Challenges},'' {\em ACM Comput. Surv.}, vol.~52, Nov.
	2019.
	
	\bibitem{10.1145/3219104.3219147}
	Z.~Sampedro, A.~Holt, and T.~Hauser, ``{Continuous Integration and Delivery for
		HPC: Using Singularity and Jenkins},'' in {\em Proceedings of the Practice
		and Experience on Advanced Research Computing}, PEARC '18, (New York, NY,
	USA), Association for Computing Machinery, 2018.
	
	\bibitem{10.5555/3294743}
	J.~Muli and A.~Okoth, {\em {Jenkins Fundamentals: Accelerate Deliverables,
			Manage Builds, and Automate Pipelines with Jenkins}}.
	\newblock Packt Publishing, 2018.
	
	\bibitem{H_b_2020}
	M.~H02b and D.~Kranzlmš¹ller, ``{Enabling EASEY Deployment of Containerized
		Applications for Future HPC Systems},'' in {\em Lecture Notes in Computer
		Science}, pp.~206--219, Springer International Publishing, 2020.
	
	\bibitem{10.1145/3332301}
	M.~Barika, S.~Garg, A.~Y. Zomaya, L.~Wang, A.~V. Moorsel, and R.~Ranjan,
	``{Orchestrating Big Data Analysis Workflows in the Cloud: Research
		Challenges, Survey, and Future Directions},'' {\em {ACM Comput. Surv.}},
	vol.~52, Sept. 2019.
	
	\bibitem{9235080}
	C.~CšŠrin, N.~Greneche, and T.~Menouer, ``{Towards Pervasive Containerization
		of HPC Job Schedulers},'' in {\em 2020 IEEE 32nd International Symposium on
		Computer Architecture and High Performance Computing (SBAC-PAD)},
	pp.~281--288, Sep. 2020.
	
	\bibitem{10.1145/3555819.3555856}
	D.~Huber, M.~Streubel, I.~Compr\'{e}s, M.~Schulz, M.~Schreiber, and
	H.~Pritchard, ``{Towards Dynamic Resource Management with MPI Sessions and
		PMIx},'' in {\em Proceedings of the 29th European MPI Users' Group Meeting},
	EuroMPI/USA'22, (New York, NY, USA), p.~57šC67, Association for Computing
	Machinery, 2022.
	
	\bibitem{greneche2022methodology}
	N.~Greneche, T.~Menouer, C.~C{\'e}rin, and O.~Richard, ``{A methodology to
		scale containerized HPC infrastructures in the Cloud},'' in {\em European
		Conference on Parallel Processing}, pp.~203--217, Springer, 2022.
	
	\bibitem{10.1145/3079856.3080246}
	N.~P. Jouppi and etc., ``{In-Datacenter Performance Analysis of a Tensor
		Processing Unit},'' in {\em Proceedings of the 44th Annual International
		Symposium on Computer Architecture}, ISCA '17, (New York, NY, USA), p.~1šC12,
	Association for Computing Machinery, 2017.
	
\end{thebibliography}
\end{document}